\documentclass[10pt]{article}

\usepackage{a4wide}
\usepackage{authblk}
\usepackage{graphicx}
\usepackage{epstopdf, epsfig}	
\usepackage{amsmath}
\usepackage{amssymb}
\usepackage[round,sort&compress,comma,authoryear]{natbib}
\usepackage{float}
\usepackage{hyperref}
\usepackage{lscape}

\title{Inertial drag-out problem : sheets and films on a rotating disc}

\author[1]{{J. John Soundar Jerome}}
\author[1]{S\'ebastien Thevenin}
\author[2]{Micka\"el Bourgoin} 
\author[1]{Jean-Philippe Matas}
\affil[1]{Univ Lyon, Universit\'{e} Claude Bernard Lyon $1$, Laboratoire de M\'{e}canique des Fluides et d'Acoustique, CNRS UMR--$5509$, Boulevard $11$ novembre $1918$, F--$69622$ Villeurbanne cedex, LYON, France}
\affil[2]{Laboratoire de Physique, ENS de Lyon, $46$ all\'ee d'Italie, F-$69364$ Lyon Cedex $7$,
France}
\begin{document}

\maketitle
\begin{abstract}
The so-called \textit{Landau-Levich-{Deryaguin}} problem treats the coating flow dynamics of a thin viscous liquid film entrained by a moving solid surface. In this context, we use a simple experimental set-up consisting of a partially-immersed rotating disc in a liquid tank to study the role of inertia, and also curvature, on liquid entrainment. Using water and UCON$^{\mbox{{\tiny TM}}}$ mixtures, we point out a rich phenomenology in the presence of strong inertia : ejection of multiple liquid sheets on the emerging side of the disc, sheet fragmentation, ligament formation and atomization of the  liquid flux entrained over the disc's rim. We focus our study on a single liquid sheet and the related average liquid flow rate entrained over a thin disc for various depth-to-radius ratio $h/R < 1$. We show that the liquid sheet is created via a ballistic mechanism as liquid is lifted out of the pool by the rotating disc. We then show that the flow rate in the entrained liquid film is controlled by both viscous and surface tension forces as in the classical \textit{Landau-Levich-{Deryaguin}} problem despite the three dimensional, non-uniform and unsteady nature of the flow, and also despite the large values of the film thickness based  flow Reynolds number.  When the characteristic Froude and Weber numbers become significant, strong inertial effects influence the entrained liquid flux over the disc at large radius-to-immersion-depth ratio, namely via entrainment by the disc's lateral walls and via a contribution to the flow rate extracted from the 3D liquid sheet itself, respectively. 
\end{abstract}

\section{Introduction}
\label{sec:intro}

\begin{figure}
\begin{center}
\epsfig{file=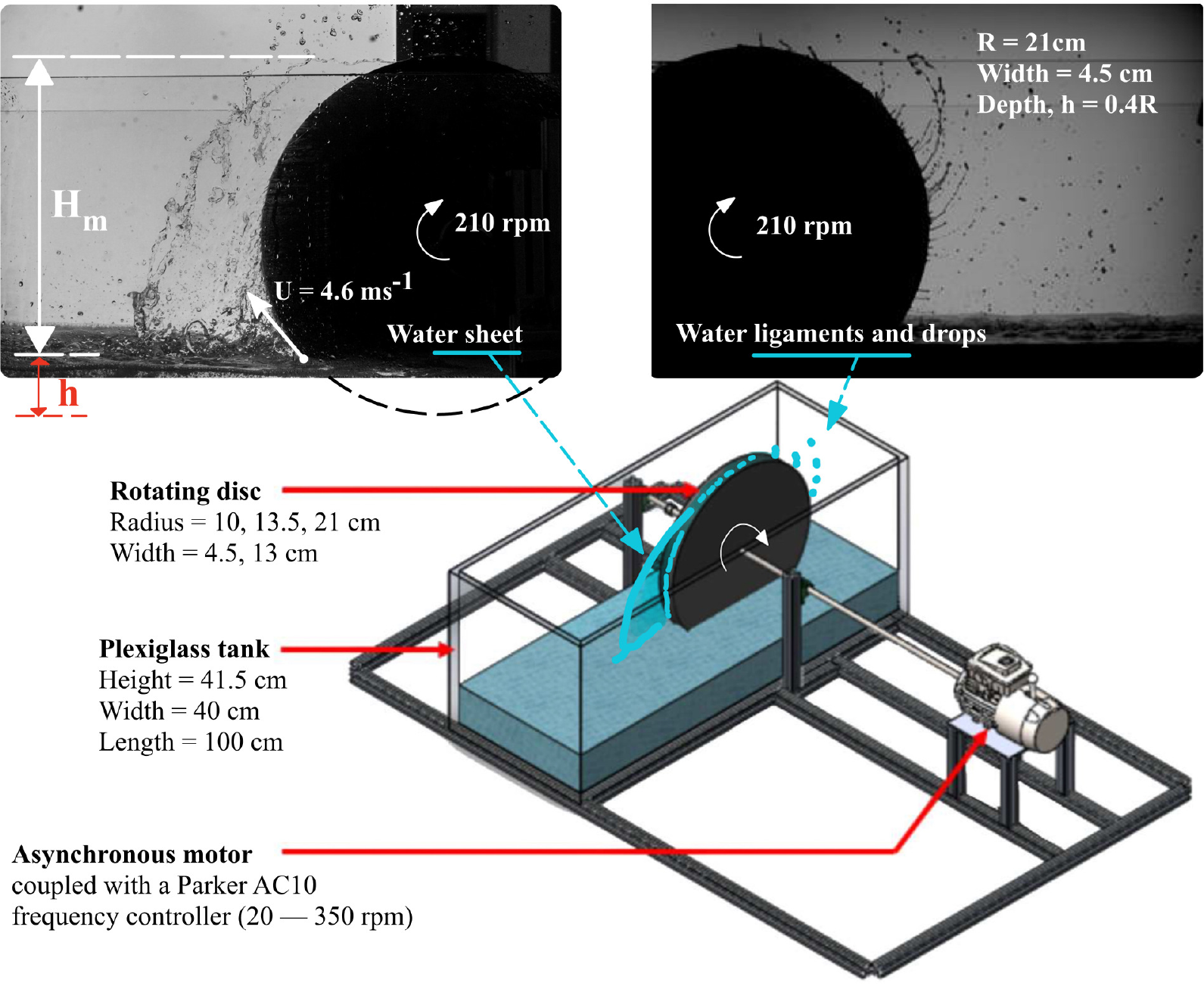, width=1\textwidth,keepaspectratio=true}
\end{center}
\caption{Water entrainment by a partially-immersed rotating disc of radius $R=21$~cm, immersion depth $h=8.4$~cm and linear velocity $U=4.6$ m~s$^{-1}$. {Two dominant liquid flow structures appear, namely, a liquid sheet and a liquid film flow which fragments into ligaments and droplets.} Schematic of the experimental set-up showing a partially-immersed rotating disc of finite width in a one meter long aquarium.}
\label{fig:setup}
\end{figure}

\begin{figure}
\begin{center}
\epsfig{file=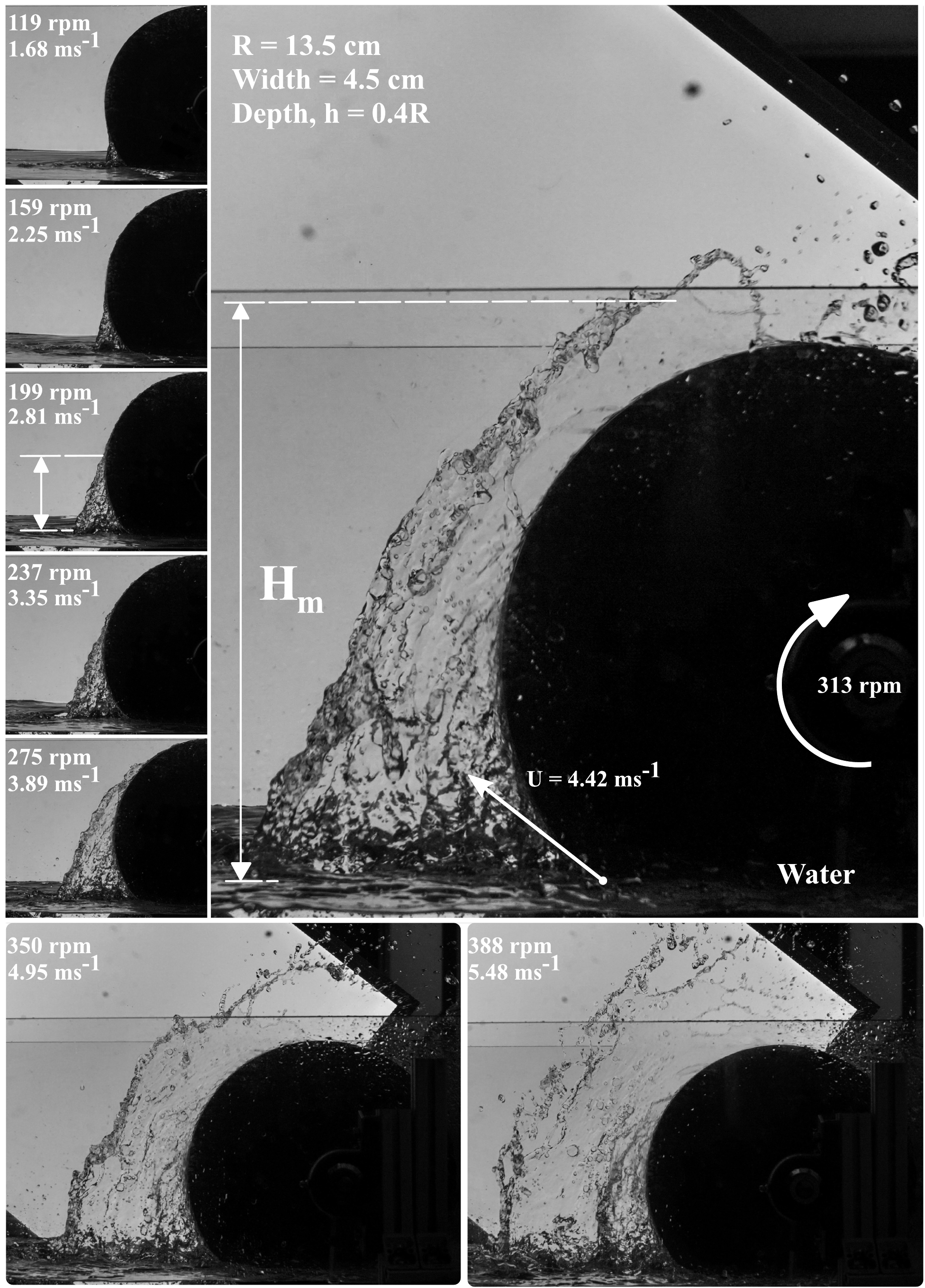, width=0.83\textwidth,keepaspectratio=true}
\end{center}
\caption{{Instantaneous side view of the inertia-driven liquid menisci on a partially-immersed rotating disc in water ($R = 13.5$ cm, ${h/R} = 0.4$).  Linear velocities $U = \Omega R$ corresponding to these conditions vary between $1.68$ m~s$^{-1}$ and $5.48$ m~s$^{-1}$. See supplementary video $1$ for more visualizations.}}
\label{fig:MenisqueEAUR13_5}
\end{figure}

\begin{figure}
\begin{center}
\epsfig{file=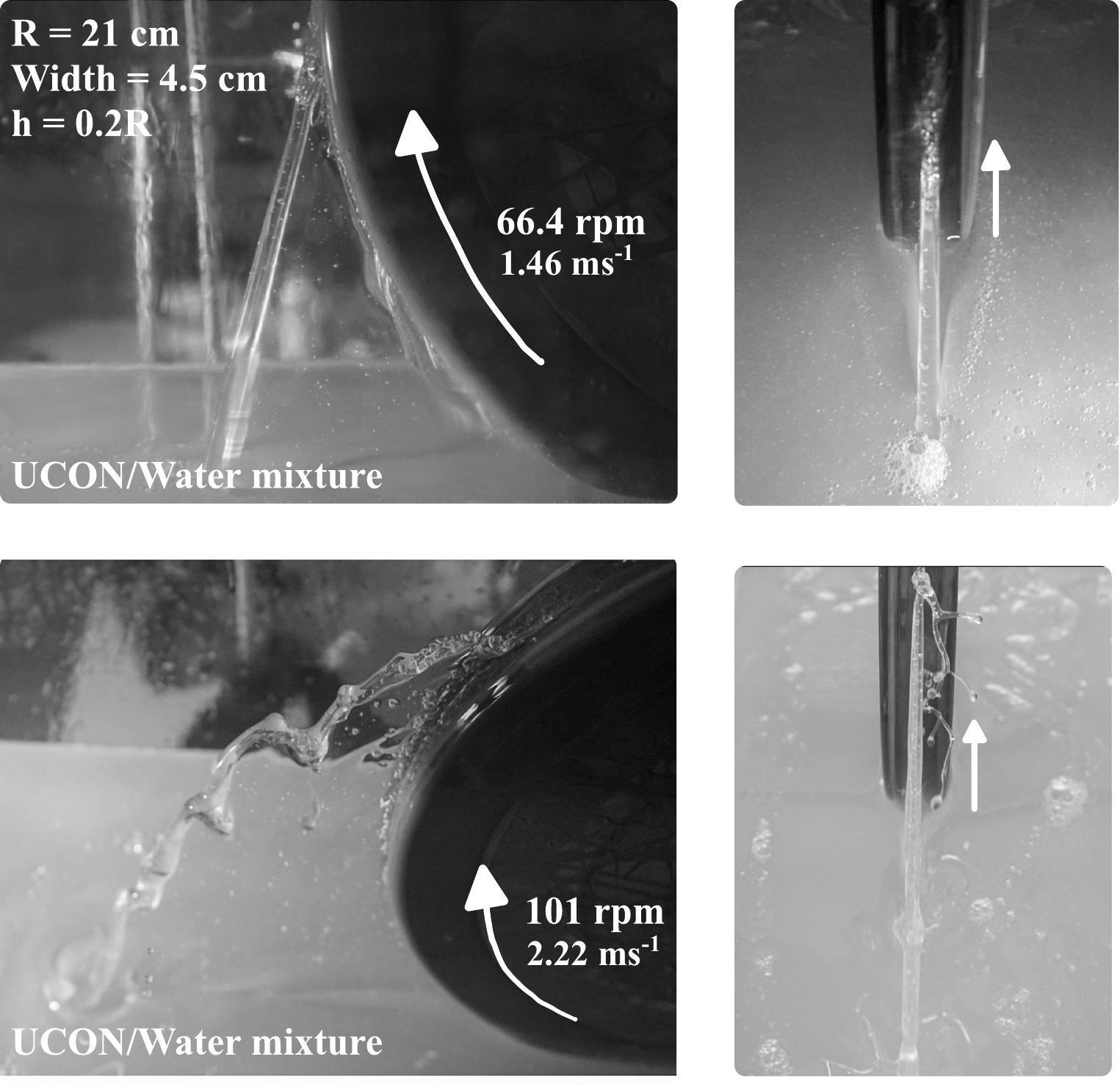,width=0.75\textwidth,keepaspectratio=true}
\end{center}
\caption{{Photographs presenting both side (left) and front (right) views of the dynamic liquid sheet in a viscous UCON/water mixtures of dynamic viscosity $\mu = 82 \times 10^{-3}$ Pa~s. Sheets can be stationary at low speeds or unsteady at higher speeds.}}
\label{fig:LiquidSheetUcon_Intro}
\end{figure}

\begin{figure}
\begin{center}
\epsfig{file=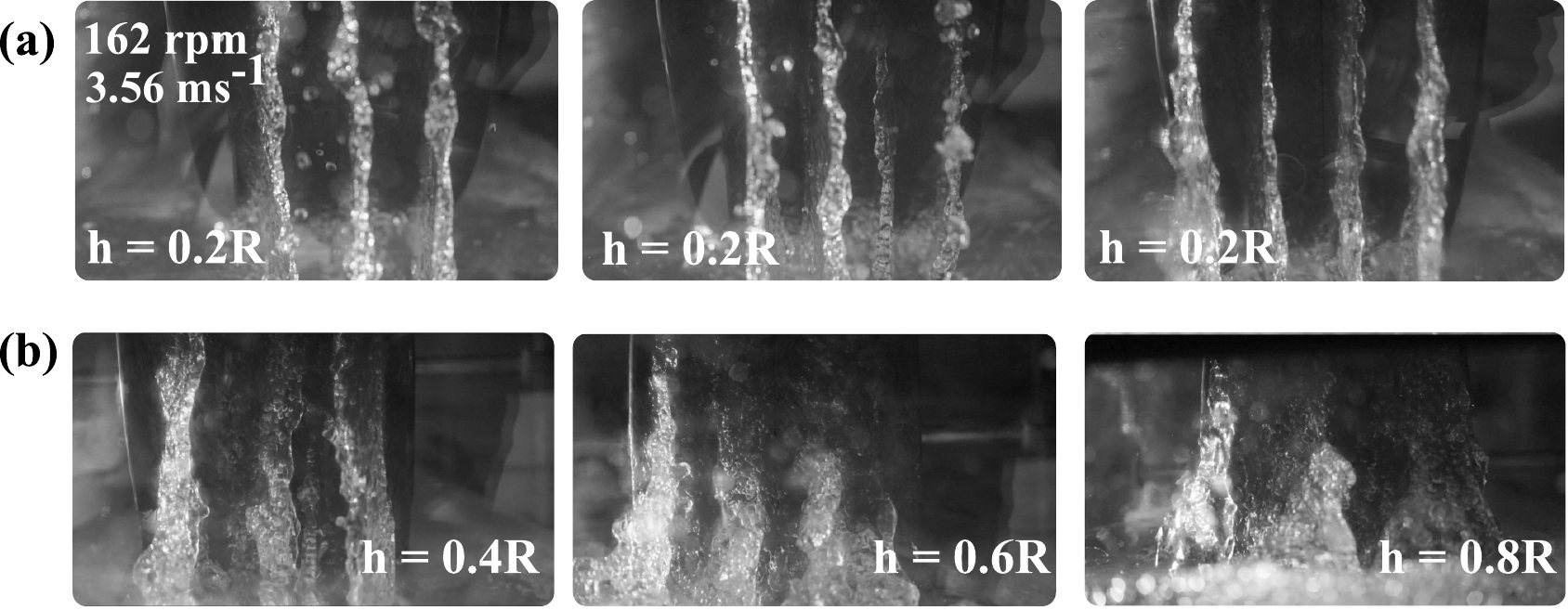,,width=0.8\textwidth,keepaspectratio=true}
\end{center}
\caption{{When a wider disc rotates in water ($R = 21$ cm, width $= 13$ cm) multiple water sheets occur: {(a) Photographs at different instants for a given wheel immersion depth $h = 0.2 R$ and $U = 3.56$ m~s$^{-1}$ showing that the number of sheets may vary with time. (b) Photographs at various immersion depths $h$ (distance between the water level and the disc's bottom.} They resemble a fully-developed inertial ribbing instability \citep{yih1960instability}.}}
\label{fig:MultipleRibs}
\end{figure}

\begin{figure}
\begin{center}
\epsfig{file=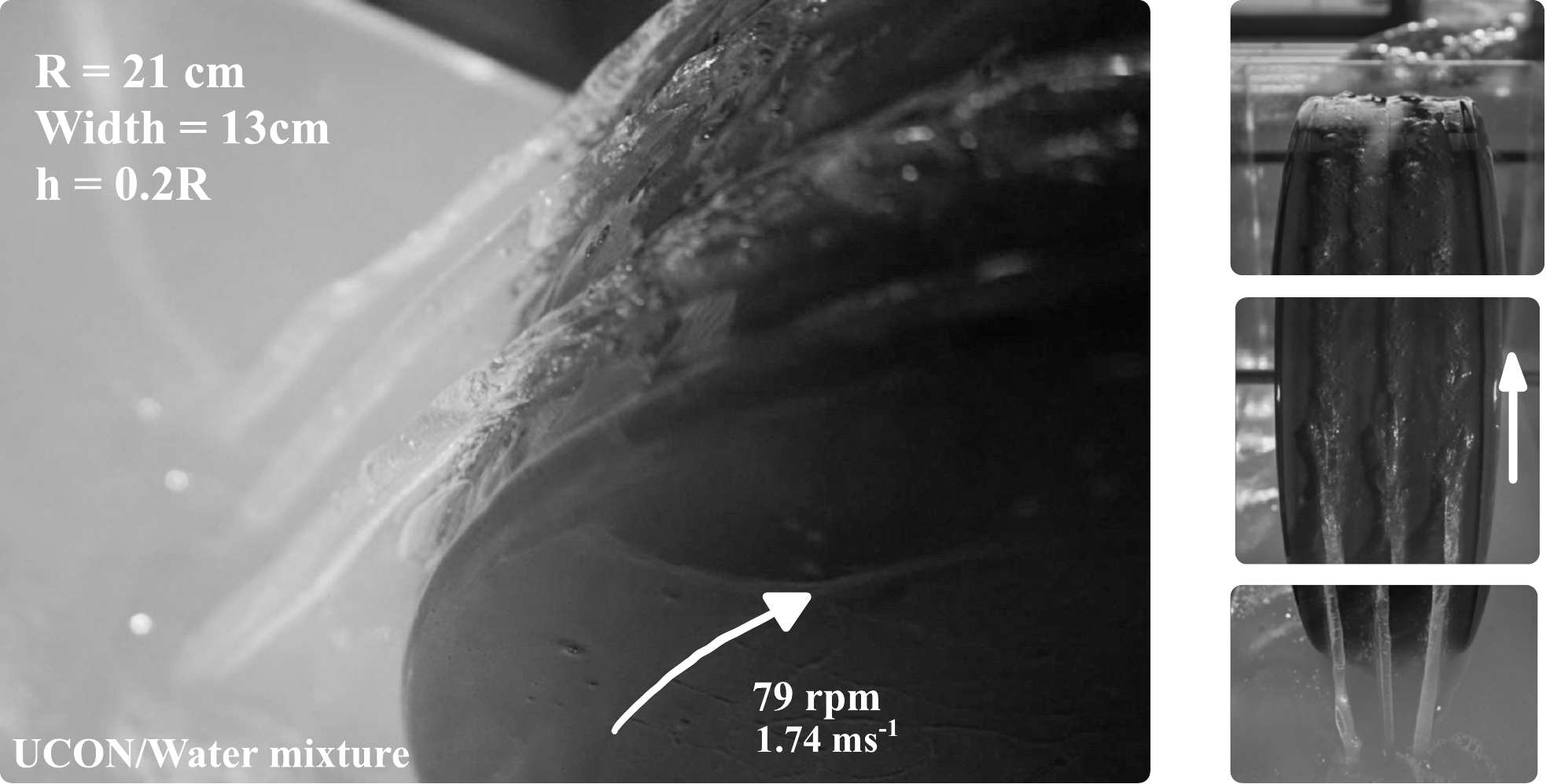,width=0.75\textwidth,keepaspectratio=true}
\end{center}
\caption{{Photographs showing multiple liquid sheets for the same UCON/water mixtures and wheel radius as in figure \ref{fig:LiquidSheetUcon_Intro} but for a wider disc (width $= 13$~cm) at $U = 1.74$ m~s$^{-1}$.}}
\label{fig:MultipleRibsUcon}
\end{figure}

{When we imagine a car rolling over a puddle, we  picture a spectacular lateral liquid splash. While this common life phenomenon is interesting on its own, the liquid flux entrained along the rotating wheel rim, its subsequent atomization, and dispersion on different parts of the vehicle, is also an important related issue. Now if we simplify both the wheel geometry and the flow domain by considering a smooth disc of finite width, partially-submerged in a rectangular liquid tank (see section \ref{sec:set-up} for details about this set-up), we can observe two global features in the resulting flow as illustrated in figure \ref{fig:setup}.} {Firstly, a liquid sheet stands out of the pool at the rear of the wheel, extending perpendicularly to its rim. And secondly, a non-uniform, unsteady liquid film is entrained on the rim, and is then atomized into ligaments and droplets. As presented in a series of photographs (see figures \ref{fig:MenisqueEAUR13_5}, \ref{fig:LiquidSheetUcon_Intro}, \ref{fig:MultipleRibs} \& \ref{fig:MultipleRibsUcon}), the liquid sheet can either be stationary at sufficiently small speeds, or  corrugated at higher speeds. When the  {disc} width is increased multiple sheets can be observed, with each of them leaving a thick trail on the rim of the wheel. These liquid sheets may even meander along the rim, and also coalesce so that the number of sheets may vary with time. In addition, air bubbles are entrained on the plunging side of the wheel.}
\begin{table}
  \begin{center}
\def~{\hphantom{0}}
  \begin{tabular}{cccccc}
      Liquid  &{Mass fraction} & Density $(\rho)$ & Viscosity $(\mu)$ &Surface tension $(\sigma)$ &Morton number\\[3pt]
         &{of UCON$^{\mbox{{\tiny TM}}}$} & kg~m$^{-3}$ & $\times 10^{-3}$ Pa~s   &$\times 10^{-3}$ N~m$^{-1}$ &$Mo = {\mu^4 g}/{\rho \sigma^3}$\\[1pt]
         \hline
       Water   						&{0} & $998 \pm 1$ &{$0.93 \pm 0.04$}  &{$72.9 \pm 0.2$ }  &{$1.9 \times 10^{-11}$}\\
       $WU1$   	&{0.09} & $1015 \pm 1$  &$5.5 \pm 0.3$  &{$54.2 \pm 0.14$}  &{$5.5 \times 10^{-8}$}\\
       $WU2$ 		&{0.13} & $1022 \pm 1$  &$11 \pm 0.5$  &{$53.7 \pm 0.14$}  &{$9.1 \times 10^{-7}$}\\
       $WU3$  	&{0.28} &$1044 \pm 1$  &$82 \pm 4 $  &{$50.7 \pm 0.04$}  &{$3.3 \times 10^{-3}$}\\
  \end{tabular}
  \caption{{Properties of working liquids in this study. Here, three different mass fraction of {UCON$^{\mbox{{\tiny TM}}}$ Lubricant $75$-H-$90$,$000$ in water (denoted by $WU$) were used. The contact angle of plexiglass-water in air and of plexiglass-$WU3$ in air varied between $29^\circ$ -- $39^\circ$ and  $32^\circ$ -- $71^\circ$, respectively. {Note that the surface tension of water is relatively high compared to bubble-treated surface-clean water samples \citep{Scott_1975}.}}}}
  \label{tab:LiquidProperties}
  \end{center}
\end{table}

{Our primary interest here is to study the simplest of the features in this inertia-dominated drag-out problem with the hope of providing some insights on high-speed liquid entrainment processes, and thereby motivating future work. For this purpose, we abandon the more {\textit{classical}} case of a long horizontal cylinder to avoid multiple sheets, and the associated multi-phase flow phenomena as well, in order to experimentally consider two simple questions : (i) how does the size of a single liquid sheet vary as a function of the disc speed $U = R \Omega$, liquid properties and immersion depth $h$, (ii)  what is the resulting entrained liquid flux on the rotating disc? {Our work is primarily experimental with an intention to bring out some scaling laws in relation to this inertia-dominated dragged-out problem.}}

{Liquid entrainment is very well documented for films on fibers that are dragged out of a liquid pool \citep{quere1999fluid}, for various coating flows between one or more horizontal rolls that feed/pull/spread a thin liquid film on a substrate \citep{ruschak1985, benjamin1995multiple, WeinsteinRuschak_AnnRevFluMech2004}, and also for film flows on the inside, or the outside, of a freely rotating drum \citep{ThoroddsenMahadevan_ExpFluids1997, seiden2011complexity}. These works are crucial to study and model thin liquid film coating flows \citep{cohen1992modern, schweizer2012liquid} and their stability \citep{yih1960instability, pitts1961flow, coyle1990stability}. In this context,  the present work may also actually be relevant to an important class of problems in both modern engineering and fundamental fluid mechanics involving high-speed entrainment of a liquid by a moving wall.}

\subsection{{LLD flows : flat plate, or fiber, withdrawal from a liquid pool}}
\label{subsec:LLDflows}
The simplest entrainment flow configuration is that of a liquid film entrained by a long vertical plate (figure \ref{fig:LLDschematics}a), or a thin vertical fiber, as it is withdrawn at a steady speed $U$ from a large reservoir of a viscous liquid of density $\rho$, viscosity $\mu$ {and surface tension $\sigma$}. It is the so-called \textit{Landau-Levich-Deryagin dip-coating flow} \citep{goucher1922thickness, morey1940thickness, levich1942dragging, derjagin1943thickness, van1958viscous}. The liquid surface tension plays a subtle role on the coating flow via the Laplace pressure at the \textit{dynamic meniscus} between the film and the reservoir. However, far away from the \textit{dynamic meniscus}, the flat entrained film is at the ambient pressure. Consequently, a pressure gradient which opposes liquid entrainment should exist along the vertical film.  {\cite{levich1942dragging} and \cite{derjagin1943thickness} were the first to remark this feature for the case when the surface tension forces dominate viscous entrainment,  \i.e. at sufficiently small capillary numbers $Ca = \mu U / \sigma$ {and at negligible inertia. They showed that the liquid film thickness far away from the reservoir is given by}}\footnote{Note that the higher-order terms in capillary number are first developed by \cite{wilson1982drag} using a matched-asymptotic {technique}.}
\begin{equation}
	\delta_f^{(LLD)} \simeq l_c \left[ 0.946 Ca^{2/3} - 0.107 Ca + \mathcal{O} \left(Ca^{4/3}\right) \right],
	\label{eqn:LLDsclaing}
\end{equation}
{where $l_c = \sqrt{{\sigma}/{\rho g}}$ is the capillary length with $g$ the acceleration due to gravity.} This pioneering result correlates well with numerous experimental investigations for small capillary numbers up to $\mathcal{O}(10^{-2})$ \citep{morey1940thickness, van1958viscous, groenveld1970low, snoeijer2008prl, maleki2011landau} and also, for moderate values of $Ca$ \citep{white1965theory, groenveld1970high, spiers1974free, Kizito1999}. 

When both inertia and surface tension are negligible compared to viscous forces and the film is flat far away from the liquid bath, the film thickness $\delta_f$ can be deduced from a simple balance between the weight of the liquid film and the viscous drag of the plate. Thus, in this \textit{viscosity-gravity driven} regime, {the relevant length scale for the film-thickness is}
\begin{equation}
	\delta_f^{(g)} = \sqrt{\dfrac{\mu U}{\rho g}} =  l_c Ca^{1/2},
	\label{eqn:DeltaPoidsVisco}
\end{equation}
{as one could expect from the classical drainage problem \citep{jeffreys_1930, Chalmers1937}. But the presence of a stagnation point in the meniscus region between the pool and the moving plate \citep{groenveld1970dip} implies that the resulting film thickness is provided by the expression (\ref{eqn:DeltaPoidsVisco}) up to a constant factor, depending on the flow situations. Subsequent experiments \citep{groenveld1970high, Kizito1999} and $2D$ computations \citep{JinAcrivosMunchPoF2005, filali2013some} suggest that such a limit is attained at $Ca \gg 1$ but only for moderate Reynolds number
\begin{eqnarray}
	Re_f = \dfrac{\rho U \delta_f^{g}}{\mu} \equiv \sqrt{\dfrac{Ca^3}{Mo}},
\label{eqn:Reynolds}
\end{eqnarray} 
based on $\delta_{f}^{g} = \sqrt{{\mu U}/{\rho g}}$ which is the relevant length scale for the film entrained by an infinite vertical wall. Note that $Re_f$ is related to the capillary number via Morton number $Mo = \mu^4 g / \rho \sigma^3$ (or, equivalently, the fluid property number) which depends only on the fluid properties at a given $g$.}

{In this context, by extending the LLD model to include inertia, \citet{deRyck1998inertia} suggested that deviations from the classical LLD limit could occur due to an increased momentum transfer from the plate at the dynamic meniscus. Experimental data of \citet[in figure $3$]{Kizito1999} indicates that deviations from the LLD law, to either larger or smaller film thicknesses, are attained at different capillary numbers depending only on the fluid property number, or Morton number $Mo$.} {While the measured film thickness for various viscous fluids plateaued out at approximately $0.7 \delta_f^{(g)}$ beyond a Weber number $Re_f Ca = We_f \approx 0.2$, it cannot be concluded from their data corresponding to less viscous liquids ($Mo \ll 1$) that this will be the case when $Re_f  \gg 1$ since the maximum Reynolds number in their experiments was only about $20$.}

{Similarly, \citet{JinAcrivosMunchPoF2005} illustrated, using $2D$ creeping flow equations ($Re_f = 0$) for the dip-coating problem, that the film flow rate per unit width attains a constant value of about $0.58 \delta_f^{(g)}U$ when $Ca \gg 1$. {They also attempted to investigate the role of inertia by varying the capillary number $Ca$ at different Morton numbers $Mo = \left\lbrace 1, 10^{-2}, 10^{-4}, 10^{-6} \right\rbrace$.} {They observed that the liquid flux dragged-out of a pool by a flat-plate is correctly predicted by the LLD scaling (\ref{eqn:LLDsclaing}), independent of $Mo$, until it reaches a maximum value {which occurs} at some characteristic capillary number {depending} only on the Morton number $Mo$. {
In particular, the entrained flow rate is smaller than $0.58 \delta_f^{(g)}U$ corresponding to their results for $Ca \gg 1$ and $Re_f = 0$ (or $Mo = \infty$). However, their numerical procedure failed beyond a critical value of capillary number {at various Morton numbers $Mo$.} Note that the maximum Reynolds number $Re_f$ attained in the $2D$ simulations of \citet{JinAcrivosMunchPoF2005} was $32$ ($Mo = 10^{-6}$) and that they, like a good number of other works \citep{middleman1978, campanella1984roll, Kizito1999, Evans2005PoF, filali2013some}, suggest the formation of \textit{cusped} menisci and \textit{wavy} free-surface structures on the liquid pool as $Re_f$ increases. }}

{} {}

{Other works on  liquid entrainment by a moving flat plate, or a fiber, in relation to LLD scaling considered the role of non-Newtonian rheology \citep{deRyck1998Langmuir}, wettability \citep{snoeijer2006prl, snoeijer2008prl}, surfactants  \citep{krechetnikov2005PoF, campana2010PoF, Mayer2012landau}, adsorbed particles on fluid interfaces \citep{campana2011PoF, dixit_homsy_2013, dong2013Langmuir, Gans_SoftMatter2019, palma_lhuissier_2019}, textured flat plates \citep{seiwert_clanet_quere_2011, Nasto2018PRF} and also, the existence of non-unique solutions for the LLD problem \citep{WeinsteinChemEngg2001, benilov_chapman_mcleod_ockendon_zubkov_2010}}.

\begin{figure}
\begin{center}
\epsfig{file=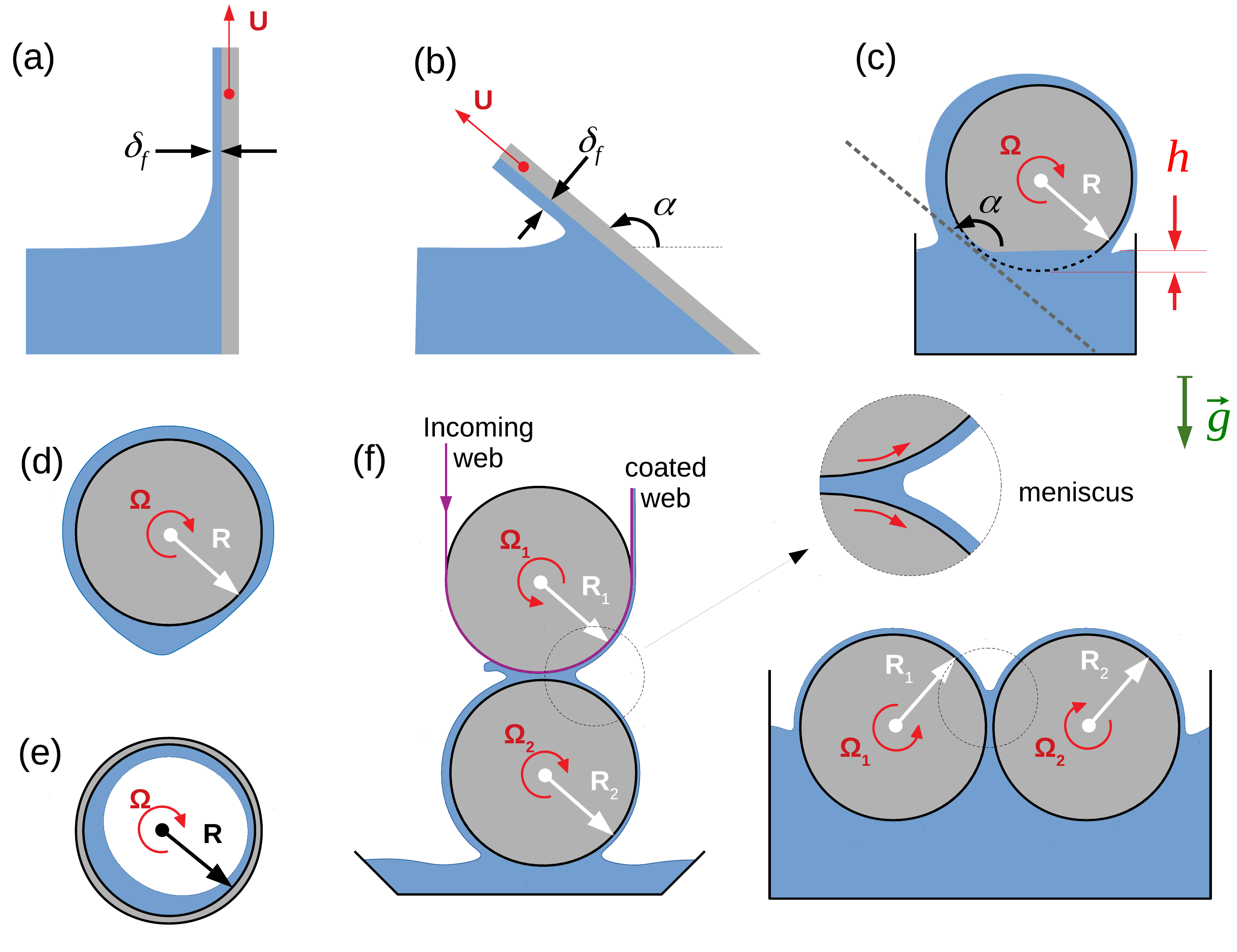,width=0.82\textwidth,keepaspectratio=true}
\end{center}
\caption{{Some common liquid entrainment configurations : (a) The classical \textit{Landau-Levich-Deryagin} (LLD) dip-coating flow. (b) \cite{tallmadge1971theory}'s variant of LLD flow which \cite{tharmalingam1978coating} used to model, as schematized in (c), the coating flow on a partially-submerged rotating drum at different immersion heights; \cite{rubashkin1967problem},  \cite{groenveld1970dip} \& \cite{middleman1978} also used this rotating cylinder set-up to explore the large capillary number limit. (d)--(e) \textit{Moffatt-Pukhnachev-Yih} (MPY) \textit{external}, or \textit{internal}, rimming flows, respectively. (f) Other examples of rotary entrainment in liquid film coating processes (for more, refer to \cite{schweizer2012liquid}). Relevant non-dimensional numbers in several previous experiments and computations are given in Table \ref{tab:ExternalCoatingData}.}}
\label{fig:LLDschematics}
\end{figure}
\subsection{{Rotary entrainment, an inclined-LLD flow?}}
\label{subsec:modifLLDflows}
{\cite{rubashkin1967problem} \& \cite{middleman1978} investigated the relevance of the LLD scaling (\ref{eqn:LLDsclaing}) for the case of a partially-immersed rotating {cylinder}. Using horizontal cylinders of radius $R = 2.5$ -- $4$ cm and about $10$ to $20$ cm long in a viscous oil bath, they reported, as others before, that the film thickness attains the limiting value of about $0.56 \sqrt{{\mu U}/{\rho g}}$ as the capillary number is increased.  They also noted,  {however briefly}, that increasing the rotational speed of the  {cylinder} destroys the \textit{dynamic meniscus}. \cite{tharmalingam1978coating} provided further film thickness measurements for a wide variety of liquids as a function of the drum's rotational speed ($\Omega$) at various {cylinder} immersion heights ($h$) under laminar flow conditions. In all these cases, there was no account of liquid sheets standing out of the liquid pool along the emerging side of the rotating cylinder similar to what is seen in photographs provided here (see figures \ref{fig:MultipleRibs} -- \ref{fig:MultipleRibsUcon}). This is probably due to the small working Reynolds number $Re_f$, or strong confinement effects.} 

These authors attempted to account for the effect of immersion height on the liquid flux entrained by a rotating drum by {modeling} the latter {as an analogous} \textit{Landau-Levich-Deryagin dip-coating flow} over an inclined flat plate (see also, \cite{tallmadge1971theory, wilson1982drag, JinAcrivosMunchPoF2005}). In particular, as depicted in figure \ref{fig:LLDschematics}b-c, if $\alpha$ is the angle between the inclined plate which is tangent to the partially immersed cylindrical drum at its line of contact with the liquid bath at rest and the horizontal line, they showed that the non-dimensional film thickness $T_0 = \delta_{f} \sqrt{\rho g \sin \alpha/\mu U}$ just after the \textit{dynamic meniscus} between the cylinder and the liquid reservoir is given by
\begin{equation}
	\dfrac{T_0}{{\left( 1 - T_0^2 \right)}^{2/3}} \sim 0.94 Ca^{1/6} {\left( \dfrac{\sin \alpha}{1 - \cos \alpha} \right)}.
	\label{eqn:LLDTharmaScaling}
\end{equation}
With the advent of numerical techniques, earlier theoretical efforts (such as \cite{soroka1971inertia} \& \cite{tharmalingam1978coating}) on the lubrication equations wherein inertial effects could be included via $1$D film flow modeling and \textit{Oseen-like} corrections were later complemented by fully non-linear, but $2$D film flow, simulations \citep{nigam1980inertia, cerro1980rapid, campanella1984roll, hassan2009cfd}. Along with \cite{tharmalingam1978coating}, these authors reported that the modified-LLD scaling for an inclined plate, as in (\ref{eqn:LLDTharmaScaling}), holds for a good range of small capillary numbers. They also noted that, by including inertial effects in the \textit{inclined-plate LLD law}, a better match was obtained with experiments when the film Reynolds number is of $\mathcal{O}(1)$. 
 
\subsection{{Rotary entrainment in roll coating and rimming flows}}
\label{subsec:Coatingflows}
{A related  problem is that of the steady and unsteady motion of a thin liquid film on the outside of a freely-rotating cylinder, or on the inside of a hollow rotating cylinder (as in figures \ref{fig:LLDschematics}d-e). They are known as \textit{external}, or \textit{internal}, rimming flows, respectively\footnote{These flows are also referred to as \textit{Moffatt-Pukhnachev-Yih} (MPY) flow \citep{yih1960instability, moffatt1977rotating, Pukhnachev1977}}. In the context of pattern formation, a large body of work exists for such thin film flows since rimming flows are prone to both azimuthal and axial non-uniformities (see \citet{Evans2005PoF, seiden2011complexity} and references therein).} {Even though  flow patterns in \textit{internal} rimming flows \citep{KovacBalmer1980_hygrocysts, ThoroddsenMahadevan_ExpFluids1997} seem to be relevant to the present study, the liquid puddle confinement is very strong in these flows. In the case of \textit{external} rimming flows, the rotating cylinder is not in continuous contact with the liquid puddle and therefore, the liquid mass entrained by the former is not constantly renewed. These differences might strongly alter the flow patterns such as the liquid sheet height and the resulting liquid entrainment process when the Reynolds number $Re_f$ is sufficiently large. In this context, questions regarding the relevant length and time scales of the liquid sheet, the entrained flux are yet not well-known.}

{Note that a very large panel of coating flows involving rollers \citep[part $3$]{schweizer2012liquid} also fall under the category of rotary entrainment. In these cases the gap between the rollers is often very small (about one-tenth of the roller radius or smaller) and the cylinder length-to-radius ratio is large. Therefore, as proposed first by G. I. Taylor, the flow field is almost always divided into two regions \citep[see chapter $12$]{schweizer2012liquid} : $2D$ flow in the liquid meniscus, or the flim-splitting region (figures \ref{fig:LLDschematics}f), and a thin film flow far away from the meniscus. In fact, these flows can also exhibit azimuthal film thickness variations \citep{ruschak1976rimming}, axial patterns \citep{pitts1961flow, coyle1990stability} and even \textit{misting}, or meniscus break-up and atomization, at high-speeds \citep{owens2011misting}. Unlike what is common in roller coaters, in the case we study here there is hardly any confinement on the emerging side of the {disc} so that the meniscus, or the liquid sheet, is not constrained by another {disc}. In addition, a long cylinder can lead to multiple liquid sheets when $Re_f \gg 1$, as already shown in figures \ref{fig:MultipleRibs} and \ref{fig:MultipleRibsUcon}.} 

{Previous experimental and computational investigations on rotary entrainment often restricted the working Reynolds number $Re_f$ based on coating thickness to small (or at best, moderate) values. When $Re_f$ was large, the meniscus was often strongly constrained as for example in roller coaters. An interested reader is referred to Table \ref{tab:ExternalCoatingData}, for more information on relevant dimensionless numbers in several earlier experimental and numerical investigations. Since both numerical and theoretical modeling of this fully three-dimensional, multi-phase and unsteady fluid flow is extremely difficult for the inertial rotary entrainment of liquids out of pool, we take a phenomenological approach in order to propose physical reasoning for the scaling laws observed in our experiment. {In this regard, we focus in the remainder of this paper on the liquid sheet formed in front of a rotating disc of finite width at $h/R < 1$, and on the related entrained liquid flux at large $Re_f$ up to $10^3$.}}

\section{Experimental set-up}
\label{sec:set-up}


As illustrated in figure \ref{fig:setup}, the experimental set-up consists of a large plexiglass tank {($40 \times 41.5 \times 100$ cm$^3$) containing the working liquid}. An asynchronous motor is used along with an AC Inverter Drive (Parker AC$10$) to {rotate} rigid PVC discs of radii $R = 10$ cm, $13.5$ cm and $21$ cm and {width} $4.5$ cm. Qualitative results for a thicker disc ($13$ cm) were also presented in the introduction but experimental data in the following sections concern only these discs of thickness $4.5$ cm. The angular velocity {can be} varied between $20$ and $350$ \textit{rpm}. {The disc is partly submerged in the liquid bath and the liquid level $h$, as depicted in figures \ref{fig:setup} and \ref{fig:LLDschematics}c, is properly verified before each run.} {The position of the wheel's axis cannot be changed in the present set-up and so various immersion depths $h$ in the range $0.05$ -- $0.8$ times the disc radius $R$ are obtained by adjusting the volume of liquid in the aquarium. 
} 

{Water and mixtures of Water/UCON$^{\mbox{{\tiny TM}}}$ oil were chosen for the experiments (table 1). The viscosity of these mixtures can be controlled readily, and they are Newtonian. Note that temperature of the set-up was never imposed externally. The ambient temperature was relatively constant, and close to $23\pm 2^\circ$C during the experiments.} Table \ref{tab:LiquidProperties} presents physical properties of the different liquids used in the experiments. Liquid density was measured using hydrometers whereas viscosity was obtained using a falling-sphere viscometer and by suitably taking into account the Reynolds numbers corrections for the viscous drag on the spheres \citep{BrownLawlerJEE2003}. {No surfactants were added to the working liquid in the present study. Surface tension measurements were performed using the pendant drop method with a tensiometer \textit{Attension Theta Flex} (\textit{Biolin Scientific AB}). {Note that, for our water samples, the measured surface tension is comparable with the value for bubble-treated surface-clean water samples \citep{Scott_1975}. This relatively high value was obtained by several independent measurements using our tensiometer.}} 

{At first, the required wheel immersion depth is reached by controlling the water level while the wheel is at rest. Then, for a given liquid a typical series of experiments is performed by running the disc at a some specific rotation rate (rpm) as indicated by a tachometer. After a few seconds, the resulting average liquid sheet height ($H_m$) and average film flow rate ($Q_f$) are measured. Then the angular velocity is increased to repeat measurements. Visualizations are carried out with the help of a LED panel (59.5 cm $\times$ 59.5 cm) placed behind the liquid reservoir. Images are acquired with a Sony $\alpha 7$ camera (CMOS Sony Exmor). The following sections describe in detail both the measurement techniques and the corresponding results obtained using this experimental set-up.}

\section{Experimental results}
\label{sec-results}

\subsection{Inertial menisci -- a liquid sheet}
\label{sec-menisci}

The series of photographs in figures (\ref{fig:MenisqueEAU}) -- (\ref{fig:Hole}) depicts the evolution of the emerging liquid sheet  as the number of rotations per minute (rpm) is increased, for a wheel radius of $21$ cm and a water level of  $h = 0.2R$. {As seen in the photographs of figure \ref{fig:MenisqueEAU}, the water level is almost constant across the lateral side of the wheel except close to the wheel's rim, at the bottom of the liquid sheet where the water level rises slightly. 
} 
At the lowest of the speeds shown here in the photographs ($40$ rpm, or $U = 0.88$~m~s$^{-1}$, corresponding to $Re_f = \rho U \delta_f^{g} / \mu \approx 270$), {a small quasi-static meniscus} appears at the contact between the wheel and the stagnant water far away from the wheel. {A close-up shows that even at this stage the flow on the disc's rim is  distinct from the classical 2D LLD flow}. Firstly, we can observe a small dimple at the location where the meniscus emerges from the liquid bath {which is reminiscent of the \textit{cusped} meniscus already observed for dip-coating flows in previous experimental \citep{Kizito1999} and numerical investigations \citep{JinAcrivosMunchPoF2005, filali2013some}.} Secondly, it is no longer a smooth two-dimensional (or axisymmetric) structure, as is the case for LLD films on flat plates (or fibres), but a closer inspection shows spatial variations in the sheet thickness and the presence of one or two liquid rims. A further increase in wheel speed leads to the formation of a liquid sheet with a single liquid rim. {At higher speeds, the sheet size increases dramatically, and the rim presents fluctuations. Holes form within the sheet, and their rupture results in drop formation (see figure \ref{fig:Hole} and supplementary video $2$) 
}
\begin{figure}
\begin{center}
\epsfig{file=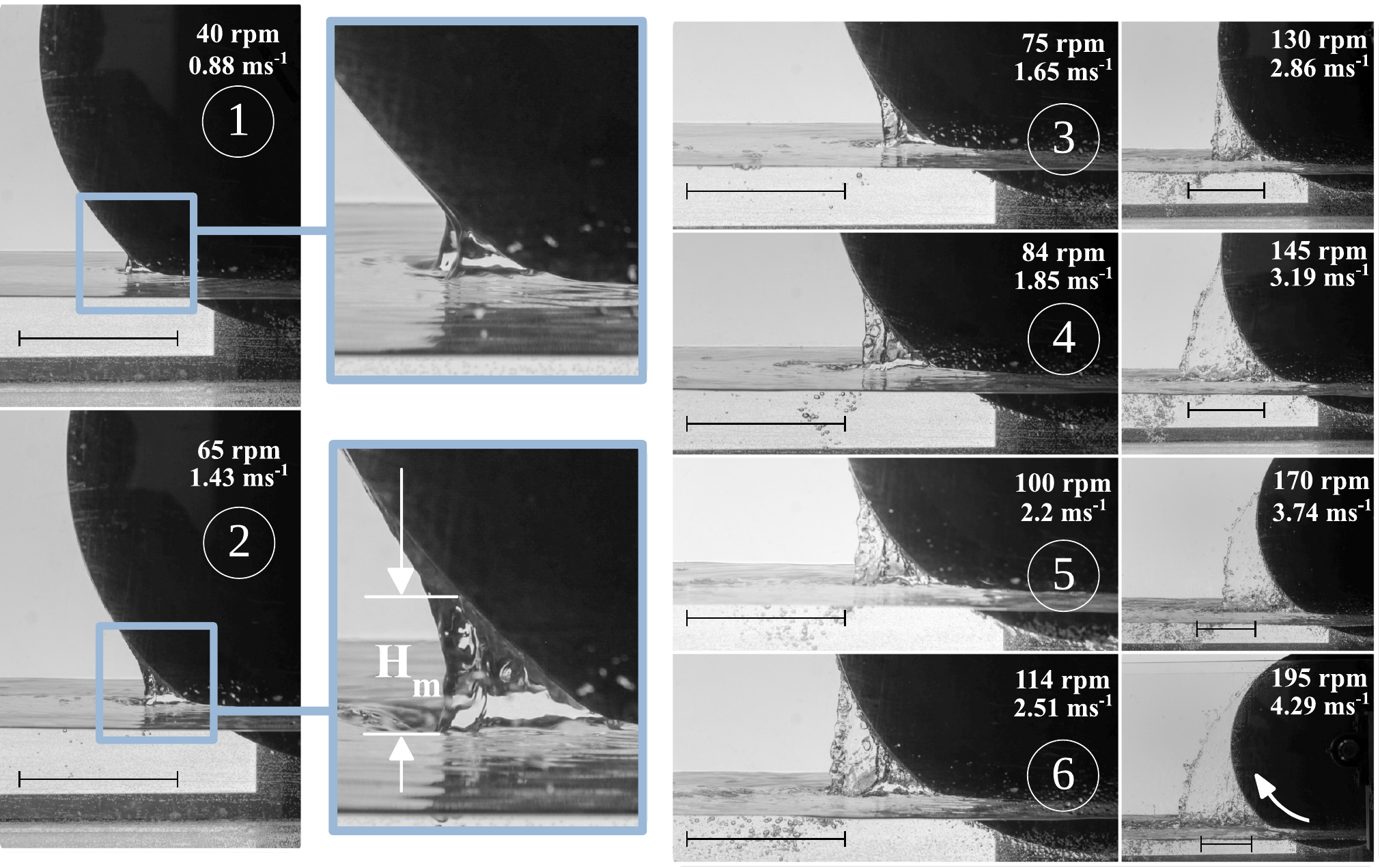,width=1\textwidth,keepaspectratio=true}
\end{center}
\caption{Instantaneous side view of the inertia-driven liquid menisci on a partially-immersed rotating disc in water ($R = 21$ cm, ${h/R} = 0.2$). Linear velocities $U = \Omega R$ corresponding to these conditions vary between $0.88$ $m$ $s^{-1}$ and $4.29$ $m$ $s^{-1}$. Displayed scales represent the half-radius of the disc $11.5$ cm.} 
\label{fig:MenisqueEAU}
\end{figure}
\begin{figure}
\begin{center}
\epsfig{file=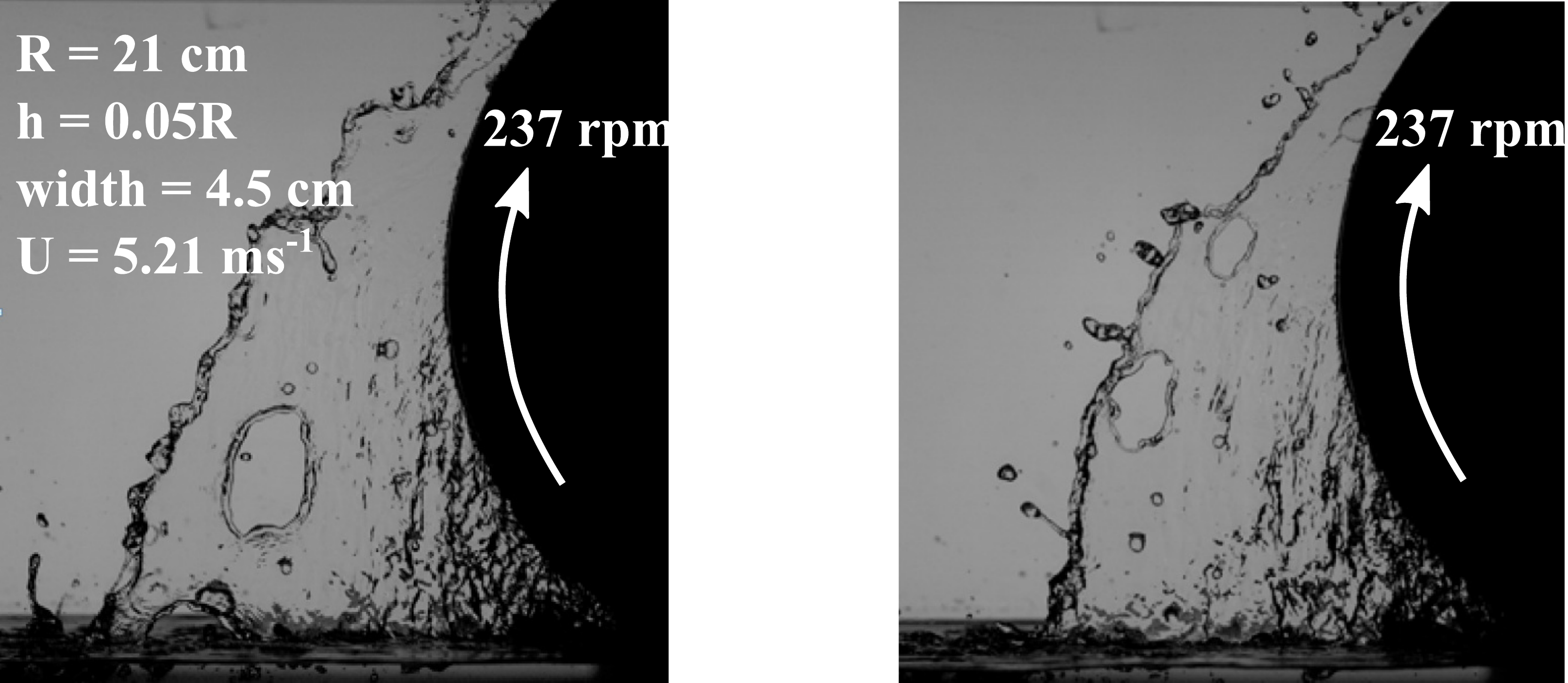,width=1\textwidth,keepaspectratio=true}
\end{center}
\caption{Visualization of the liquid sheet entrained by the wheel, for $R=21$~cm, $h=1$~cm and $U=5.21$~m/s: holes form regularly on the water sheet and disintegrate it, leading to the formation of droplets.}
\label{fig:Hole}
\end{figure}

\begin{figure}
\begin{center}
\epsfig{file=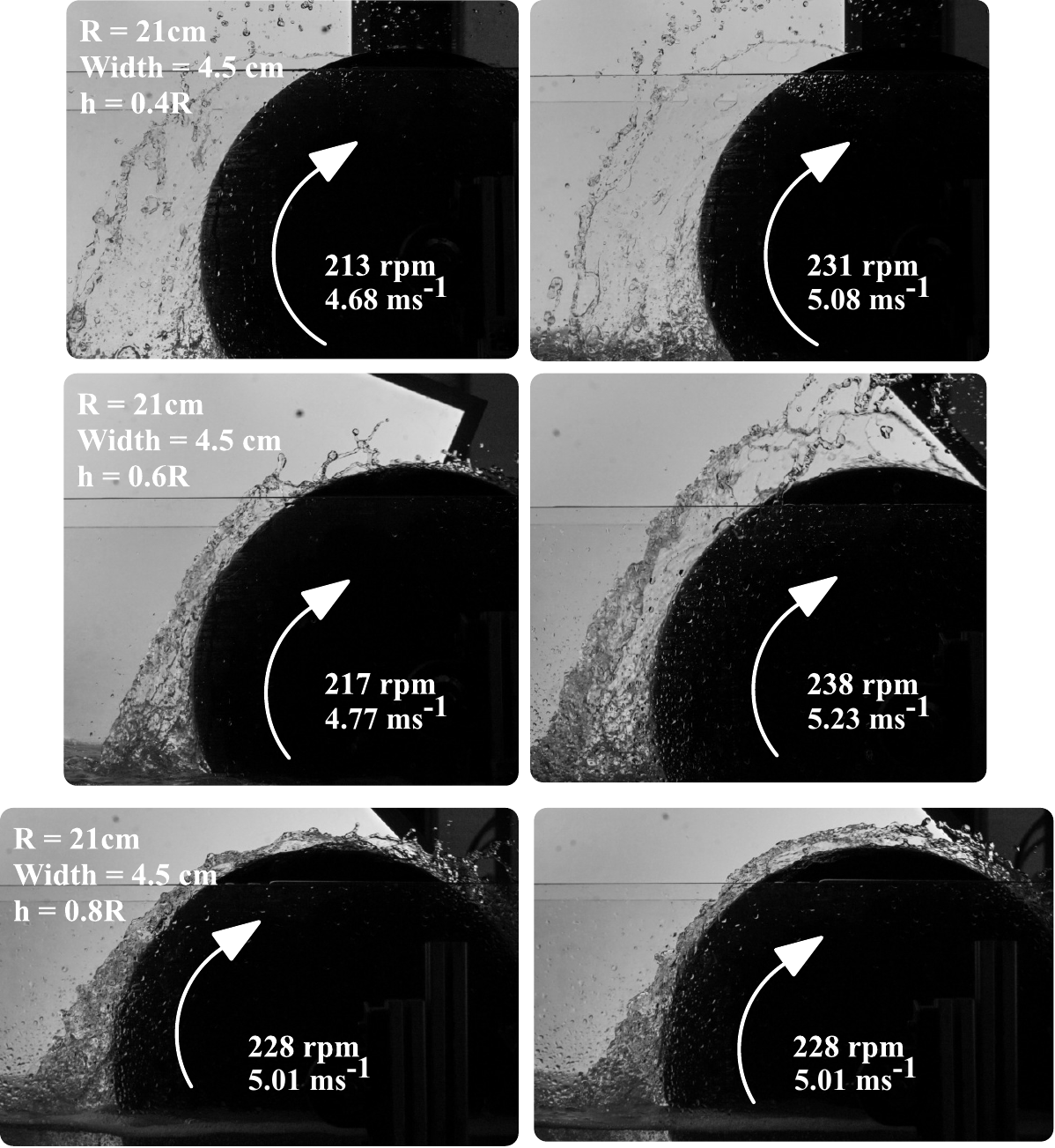,width=0.8\textwidth,keepaspectratio=true}
\end{center}
\caption{{Instantaneous side view of the inertia-driven liquid menisci on a partially-immersed rotating disc in water ($R = 21$ cm) at various immersion depth ${h/R}$. Linear velocities $U = \Omega R$ corresponding to these conditions vary between $4.68$ m~s$^{-1}$ and $5.23$ m~s$^{-1}$. {The bottom two images compare the sheet height and the related liquid entrainment at the same speed but at different times.} 
}}	
\label{fig:MenisqueEAU_R21}
\end{figure}

\begin{figure}
\begin{center}
\epsfig{file=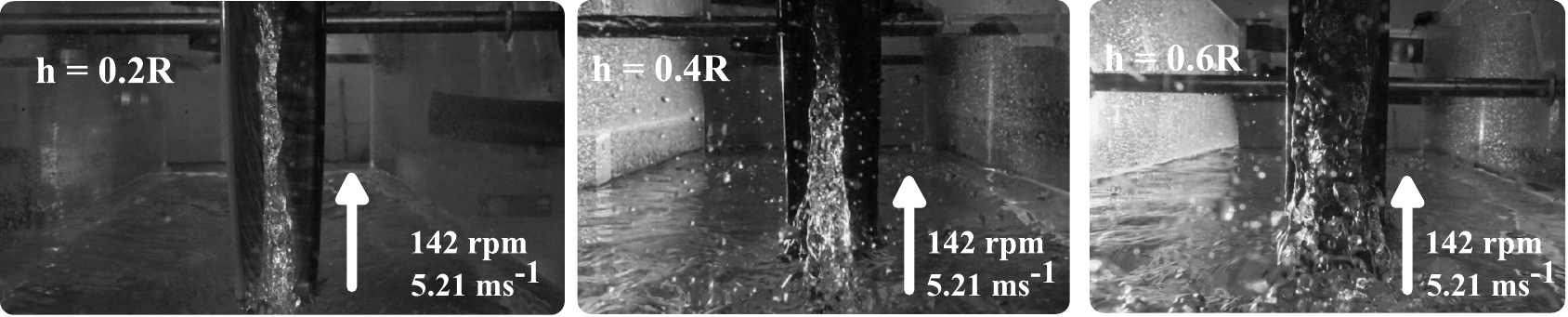,,width=0.8\textwidth,keepaspectratio=true}
\end{center}
\caption{{Instantaneous front view of the inertia-driven liquid menisci on a partially-immersed rotating disc in water ($R = 21$ cm) at various immersion depths ${h/R}$ but for the same speed $U = 3.12$ m~s$^{-1}$. {A sheet as thick as the wheel width  is obtained for the case of $h/R = 0.8$.} 
}}
\label{fig:MenisqueEAU_R21_DepthEffect}
\end{figure}

\begin{figure}
\begin{center}
\epsfig{file=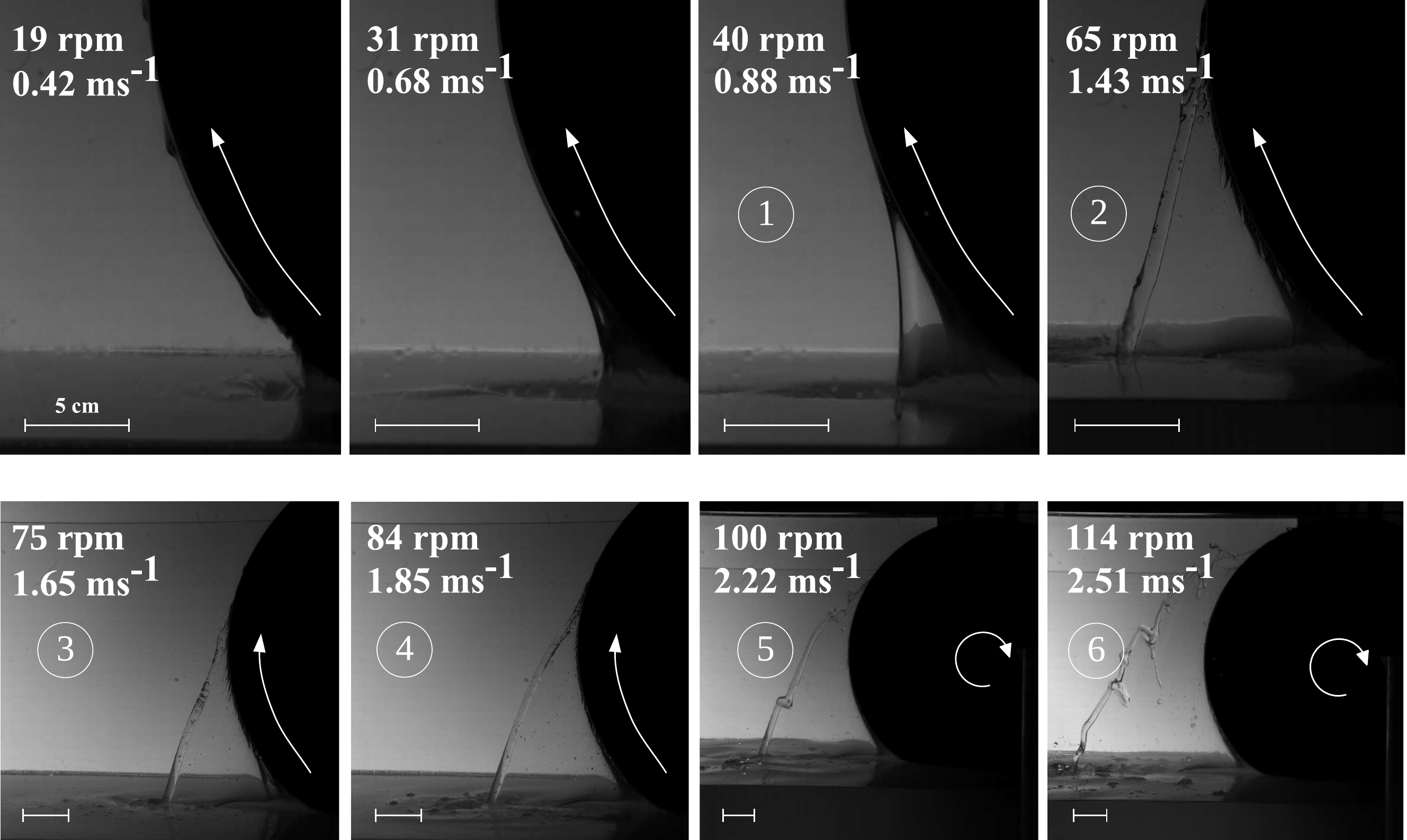,width=1\textwidth,keepaspectratio=true}
\end{center}
\caption{{Instantaneous side view of the inertia-driven liquid menisci on a partially-immersed rotating disc ($R = 21$ cm, ${h/R} = 0.2$) in a viscous liquid ($\mu = 82$ mPa s), a mixture of UCON$^{\mbox{{\tiny TM}}}$ oil and water. Linear velocities $U = \Omega R$ corresponding to these conditions vary between $0.42$ m~s$^{-1}$ and $2.51$ m~s$^{-1}$.  For a video see supplementary video $4$ to have closer look at the sheet, the rim and parabolic trajectories of entrained small bubbles inside the sheet when the wheel is rotating at $65$ rpm.}} 
\label{fig:MenisqueUCON}
\end{figure}
{The influence of the immersion depth $h$ of the wheel is illustrated in figures \ref{fig:MenisqueEAU_R21} \& \ref{fig:MenisqueEAU_R21_DepthEffect}. In the former, the first four photographs (top and middle) compare instantaneous side views of the water sheet over a rotating disc of same radius $(R = 21$ cm$)$ and approximately same rotation speeds but with different liquid depths, namely, $h/R = 0.4$ \& $h/R = 0.6$. Firstly, it is clearly visible that the water sheet climbs as high as the wheel at about $215$ rpm, or even higher at faster rotation rates. Secondly, the water depth $h$ does not seem to influence very much the water sheet height. A careful observation indicates that the water sheet is thicker at its base, just at the emerging side of the wheel's rim, and thereafter the sheet thins out. Observing the last of the two images in figure \ref{fig:MenisqueEAU_R21}, it is inferred that the effect of water depth on the sheet height is not negligible for the case of $h/R = 0.8$ as compared to smaller depths. Moreover, instead of a thin liquid sheet a very thick mass of liquid is ejected in front of the wheel and also, a relatively thicker layer of water is entrained on top the wheel. Front views of these configurations, as presented in figure \ref{fig:MenisqueEAU_R21_DepthEffect} at smaller speeds, also illustrate these observations on the sheet height and sheet thickness.}

The same phenomenology, with a few exceptions, is observed when a more viscous liquid is used (figure \ref{fig:MenisqueUCON}). Here, the liquid used is a Water-UCON$^{\mbox{{\tiny TM}}}$ oil mixture $WU3$ (see Table \ref{tab:LiquidProperties} for physical properties). Already at $U = 0.42$~m~s$^{-1}$ {($Re_f \approx 10$ and $Ca \approx 0.7$)}, a three dimensional quasi-static meniscus appears. In addition, pendant drops are also seen to descend along the rim of the rotating disc (see supplementary video $3$). This film flow is similar to that observed by \cite{Evans2004PoF, Evans2005PoF} in their numerical simulations of \textit{external} rimming flows, or the classical \textit{Moffatt-Pukhnachev-Yih} (MPY) flow, at low speeds. No trace of descending pendant drops is seen at larger speeds but a quasi-steady liquid sheet emerging from the liquid bath is clearly visible. {These images suggest that even in this highly viscous liquid, a lubrication approximation for the related flow {may be} irrelevant.} At speeds up to $1.85$~m~s$^{-1}$ {($Re_f \approx 90$)}, a fine stable liquid sheet which terminates with a thick rim flow is observed. From these images of Water-UCON$^{\mbox{{\tiny TM}}}$ oil mixtures ($WU3$), it is evident that the liquid rim is a result of a capillary recession, as in a Savart sheet \citep{savart1833memoire, villermaux2013viscous}. This  recession is known to occur at the Taylor-Culick velocity of $v_c=\sqrt{\sigma/\rho \delta_{s}}$ where $\delta_{s}$ is the local liquid sheet thickness \citep{TaylorCulick_1959, TaylorCulick_1960, SavvaBush_JFM2009}. As the speed is further increased, the rim of the sheet shows strong corrugations (see also, figure \ref{fig:LiquidSheetUcon_Intro}). {Numbers are indicated for several images in figure \ref{fig:MenisqueEAU} (for water) and figure \ref{fig:MenisqueUCON} (for UCON$^{\mbox{{\tiny TM}}}$). They correspond to identical rotation rates for these series of experiments, and are provided to help the reader compare the height of the liquid sheet for different fluids. Such a comparison illustrates that the sheet height $H_m$ measured from the surface of the pool is about $3$ to $5$ times larger for the case of the most viscous liquid ($WU3$) than for the water case, at a given disc speed $U$.}

\begin{figure}
\begin{center}
\epsfig{file=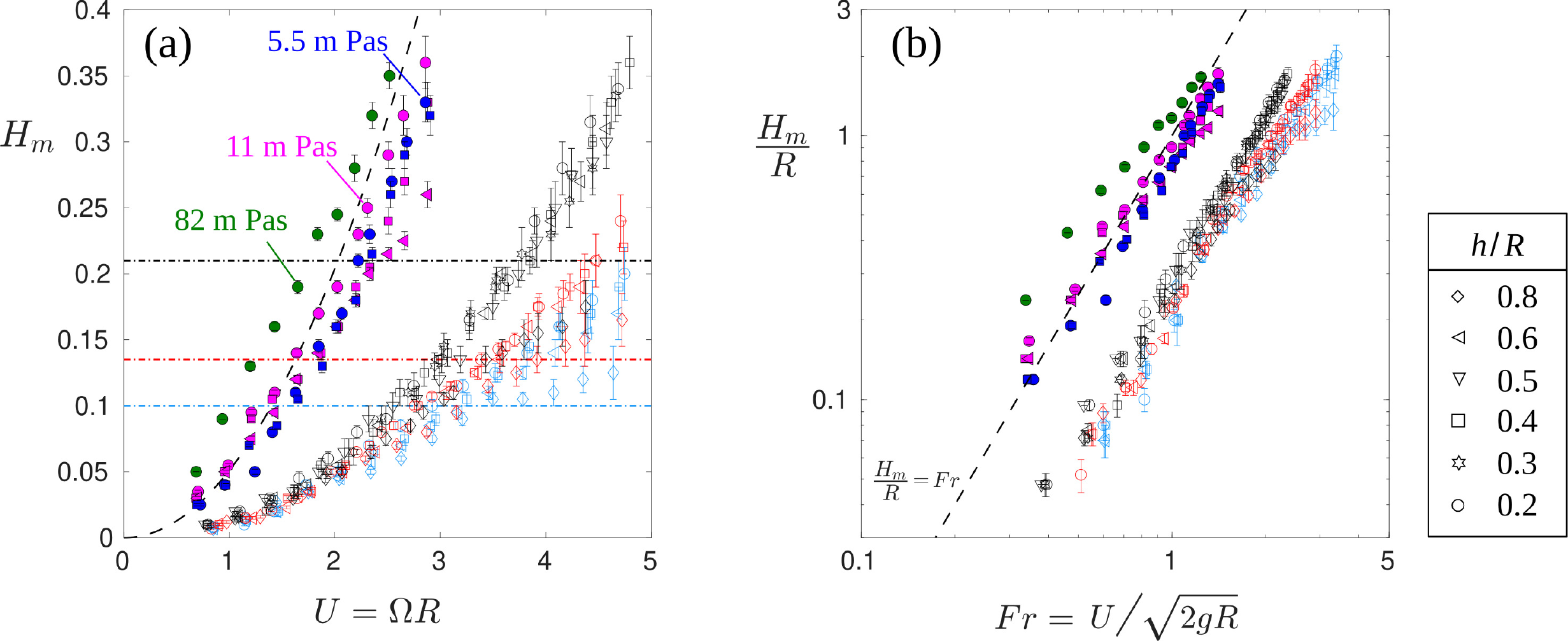,width=1\textwidth,keepaspectratio=true}
\end{center}	
\caption{{Average liquid sheet height $H_m$ (in meters) with respect to the horizontal liquid level in the tank as shown in figure \ref{fig:setup} -- \ref{fig:MenisqueEAUR13_5}. (a) $H_m$ Vs rotational speed $U = R \Omega$ for various disc radii $R = 10$ ({blue}), $13.5$ (red), $21$ (black) cm and immersion depth to radius ratio $h/R$. Data correspond to the experiments with both pure water (open symbols) and various UCON$^{\mbox{{\tiny TM}}}$/Water mixtures ($R = 21$ cm, filled symbols). {Dash-dot} line correspond to radius of the wheel. (b) Same data rescaled with respect to two different Froude number scaling. Dashed lines represent $H_m = U^2/2g$ (or simply, $H_m/R = Fr^2$). Note that the error bars indicate the amplitude of the fluctuations in the instantaneous value of sheet height.}}
\label{fig:Menisque}
\end{figure}


\subsubsection{{Time-averaged sheet height}}
{We have carried out measurements of the mean \textit{maximum} liquid sheet height $H_m$ above the surface of the pool, measured by visualization over a duration of 10s. As seen in figures \ref{fig:setup} -- \ref{fig:MenisqueEAUR13_5} (and also, {\ref{fig:MenisqueEAU} \& \ref{fig:MenisqueUCON}}), $H_m$  is not necessarily measured at the point where the liquid sheet meets the wheel. These results are presented in figure \ref{fig:Menisque} wherein the error bars correspond to minimum/maximum of the sheet height during the measurement window.} Figure \ref{fig:Menisque}(a) presents these measurements as a function of the rotation speed $U$ for three disc radii and several depths $h$ for all working liquids. Here, filled symbols correspond to viscous UCON$^{\mbox{{\tiny TM}}}$/Water mixtures and open symbols correspond to \textit{pure} water. 

{Consider now all data when the working liquid is water. The sheet height $H_m$ increases monotonically with velocity. In figure \ref{fig:Menisque}(a), each symbol corresponds to a different depth of immersion $h$ of the rotating wheel whereas colours of open symbols denote wheel radius $R$. We observe that the depth-to-radius ratio $h/R$ and the wheel radius $R$ seem to have little influence on the average height of the liquid sheet $H_m$, except for the cases at larger velocities. Especially for velocities about $3$~m~s$^{-1}$ and higher, the role of wheel radius is not negligible. We attribute this effect to the wheel's finite-size effect arising from the fact that here $h + H_m$ is $\mathcal{O}(R)$, or greater. So, it is expected that a smaller wheel (larger curvature) will lead to a smaller sheet for a given linear velocity $U$. In addition, this effect seems to be much more significant when $h/R = 0.8$ (denoted by $\diamond$) at all wheel radii and also, when $h/R = 0.6$ (denoted by $ \triangleleft$) for the smallest of the wheels. As already discussed in the previous section in relation to figure \ref{fig:MenisqueEAU_R21_DepthEffect}, the effect of the wheel immersion $h$ is to increase the sheet thickness. In particular, when $h/R = 0.8$ the sheet thickness is as large as the wheel's width and so, it is distinctly different from the cases when $h/R$ is smaller. Furthermore, as the velocity is further increased, the sheet exhibits large fluctuations and strong fragmentation which is readily evident from increasing length of the errorbars in figure \ref{fig:Menisque}(a). At much higher velocities, the top of the liquid sheet stretches even beyond the top of the rotating disc (see figures \ref{fig:MenisqueEAUR13_5} and \ref{fig:MenisqueEAU_R21} for example).}%

{Filled symbols in figure \ref{fig:Menisque}(a) denote data for liquids with different mass fractions of UCON$^{\mbox{{\tiny TM}}}$ oil in water. The time-averaged liquid sheet height $H_m$ exhibits a similar trend with speed $U$, as for the cases of \textit{pure} water. In addition, $H_m$ depends on the viscosity of water and UCON$^{\mbox{{\tiny TM}}}$ oil mixture of varying viscosity ($\mu = 5.5, 11$ and  $82$  $\times 10^{-3}$Pa.s). By comparing with the case of water (open symbols), it can be concluded that a strong increase in sheet height $H_m$ occurs when viscosity is increased. Besides, for the highest viscosity liquid, the sheet height seems to increase linearly with the wheel speed.}


A simple picture to describe the formation of the liquid sheet can be obtained by considering the motion of liquid particles as \textit{ballistic} when they are dragged out of the bath by the rotating drum at a velocity proportional to the wheel speed $U = \Omega R$. The fluid particles then follow a \textit{free-flight} trajectory in the liquid sheet and attain a height $H_m \propto U^2/2g$. Irrespective of the working liquid, figure \ref{fig:Menisque}(b) illustrates that $H_m$ is indeed proportional to $U^2/2g$ when all data from figure \ref{fig:Menisque}(a) are non-dimensionalized in terms of $H_m/R$ versus Froude number $Fr = U / \sqrt{2g R}$ based on the disc radius. 
This suggests that the simple \textit{ballistic} model captures the overall trend and provides a first approximation for the {time-averaged} inertial sheet height. 

{}

{Nonetheless, data in Figure \ref{fig:Menisque}(b) suggests that $H_m/R \sim Fr$ for all liquids used in our experiments with an $\mathcal{O}(1)$ factor that depends on the liquid viscosity. In fact, }a closer look at the data makes it clear that $H_m$ is significantly smaller than this ballistic height as $2 g H_m/U^2 \sim 0.3$, for the case of water. In addition, this ratio seems to decrease at larger speeds. Firstly, the fluid elements which quit the rim before ending up in the sheet do not all have the same momentum and so, the corresponding fluid element may only reach a smaller fraction of the ballistic height $U^2/2g$. {Secondly, capillary recession is expected to truncate the liquid sheet when it stretches and thins out, as evidenced by the formation of the liquid rim bordering the liquid sheet on its outer perimeter. Therefore, increasing viscosity is expected to delay this capillary recession, and lead to larger sheets, in accordance with the observed trend.}

\begin{figure}
\begin{center}
\epsfig{file=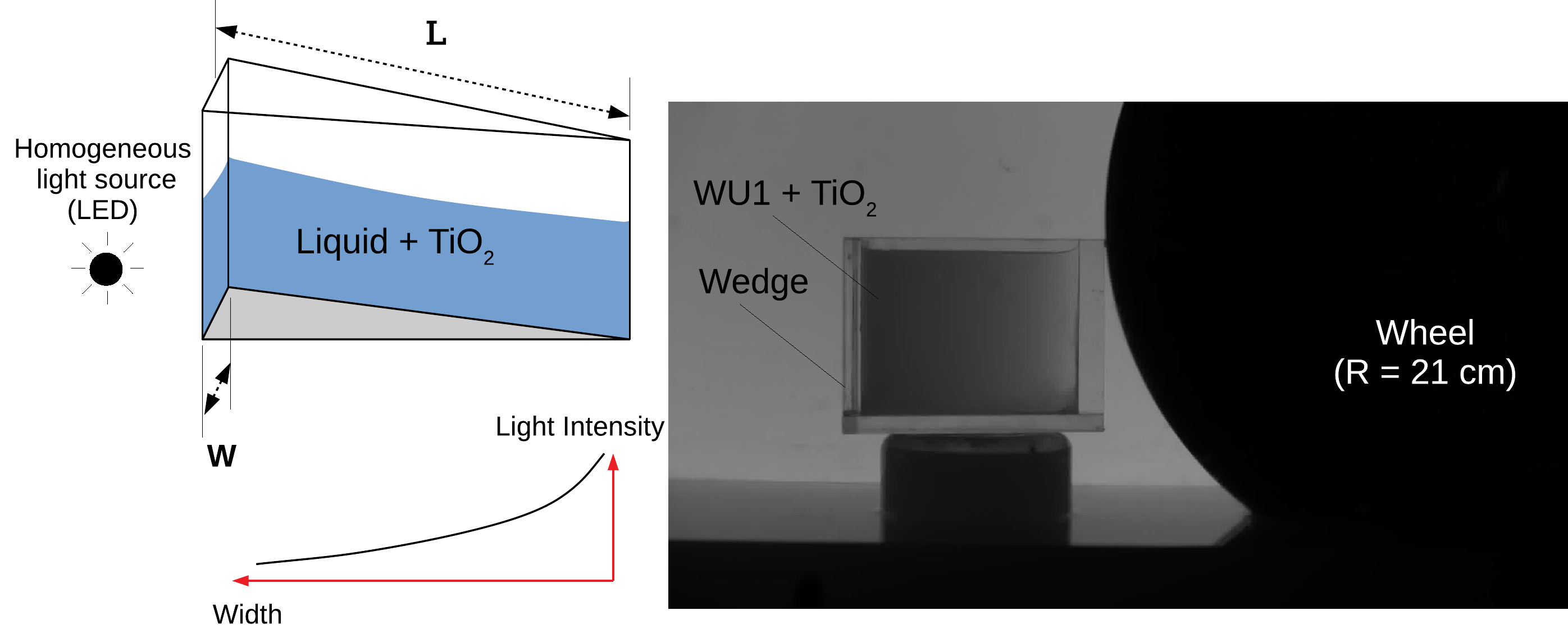,width=1\textwidth,keepaspectratio=true}
\end{center}
\caption{{Schematic and photographs showing the wedge used for the calibration of liquid sheet thickness. Two wedges with different lengths and widths were used : (1) $L = 12.5$ cm \& $W = 4.5$ cm ; (2) $L = 14$ cm \& $W = 1$ cm. Minimum measurable thickness is $2$mm.}}
\label{fig:ThicknessCalibration}
\end{figure}

\subsubsection{Time-averaged liquid sheet thickness}
In order to better understand the effect of viscosity on the retardation of the Taylor-Culick cut-off and thereby, explain the larger heights observed when viscosity is increased, further information on the inertial liquid sheet thickness is obtained experimentally. 
{A standard technique is to mix the working liquid with a small amount of fine titanium oxide particles (Ti O$_2$) whereby the liquid opacity increases. For calibration, small wedge-shaped plexiglass tanks 
are used (figure \ref{fig:ThicknessCalibration}).  When these tanks are filled with opaque liquid/Ti O$_2$ mixtures and illuminated with the back light LED panel, the intensity of the light passing through the tank varies depending on the local width. This technique is used to properly calibrate the relationship between liquid thickness and the light intensity received by the camera. It is then possible to deduce the local thickness of the sheet for various wheel immersion depths and rotational speeds. The profiles and contour plots of the \textit{mean} sheet thickness $\delta_s$ are deduced by averaging over $1500$ images acquired during $30$ seconds each and then applying this calibration. Data for these time-averaged sheet thickness profiles are depicted in figures \ref{fig:MenisqueEpaisseur} and \ref{fig:MenisqueEpaisseurComp}. Note that the maximum and the minimum measureable thicknesses are about $4.5$ cm and $2$ mm, respectively.}

\begin{figure}
\begin{center}
\epsfig{file=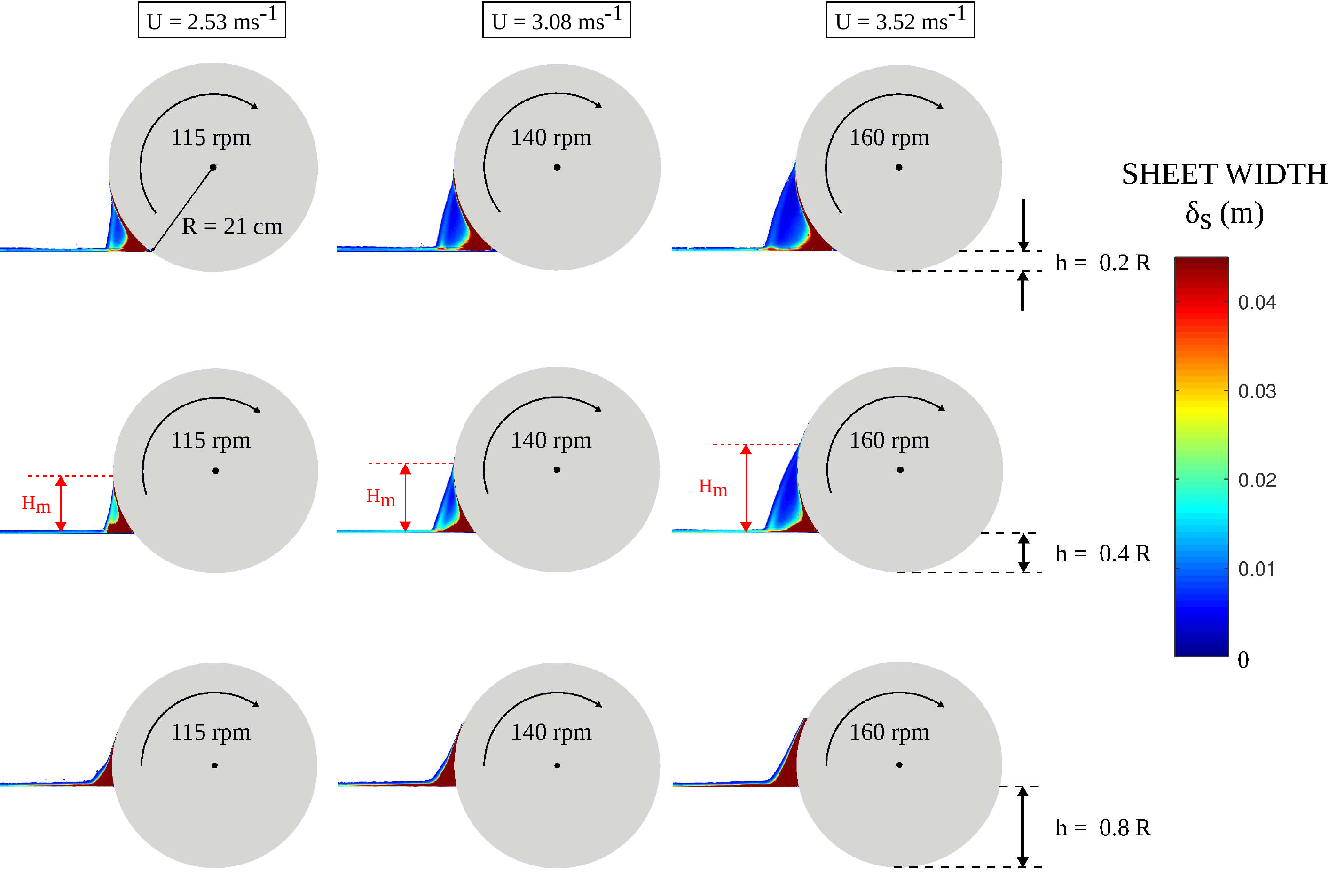,width=1\textwidth,keepaspectratio=true}
\end{center}
\caption{{Liquid sheet thickness ($\delta_s$) at different immersion depth to radius ratio ${h/R}$ and rotational speeds for water. All data correspond to a wheel radius of $21$~cm. Maximum measurable thickness is $4.5$~cm (width of the drum), minimum measurable thickness is $2$~mm.}}
\label{fig:MenisqueEpaisseur}
\end{figure}
\begin{figure}
\begin{center}
\epsfig{file=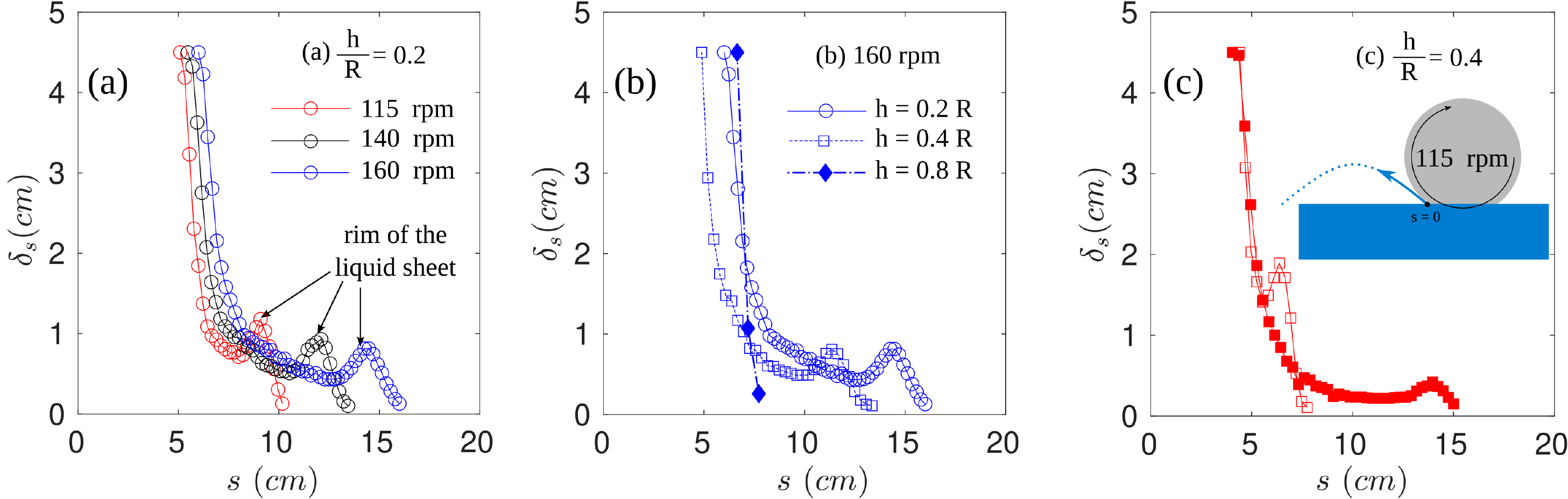,width=1\textwidth,keepaspectratio=true}
\end{center}
\caption{Sheet thickness $\delta_s$ as a function of the distance $s$ traveled along a ballistic trajectory by a hypothetical fluid element that is ejected with an initial velocity equal to $U = R \Omega$ at the emerging side of the wheel ($R = 21$cm). Here, $s = 0$ is at the line of contact between the disc and the horizontal liquid surface. }
\label{fig:MenisqueEpaisseurComp}
\end{figure}

For both $h/R = 0.2$ and  $h/R = 0.4$, the contour plots in figure \ref{fig:MenisqueEpaisseur} illustrate that the sheet thickness $\delta_{s}$ is not constant. It is thicker than the wheel's rim at its base where fluid elements are ejected with a kinetic energy proportional to $U^2/2$, and it rapidly thins out to attain a critical thickness at which capillary recession occurs. This recession then leads to the formation of a thick liquid rim, clearly visible on the thickness profiles. As mentioned previously, this recession leads to sheet truncation at a location where the local fluid velocity is smaller than the Taylor-Culick velocity of $v_c=\sqrt{\sigma/\rho \delta_{s}}$ \citep{TaylorCulick_1959, TaylorCulick_1960, SavvaBush_JFM2009}. Thus, for a given water depth, the critical sheet thickness where the sheet forms a thick rim should decrease with increasing rotational speed.  This is indeed observed in figure \ref{fig:MenisqueEpaisseurComp}(a) {wherein $\delta_s$ attains a minimum of about $2$ mm and $4$ mm, respectively, as the rotation rate increases from $115$ to $160$ rpm. On the other hand, at a given rotation rate $160$ rpm for both $h/R = 0.2$ and $h/R = 0.4$, figure \ref{fig:MenisqueEpaisseurComp}(b) indicates that $\delta_s$ attains a minimum of about the same value $4$ mm. So, the depth-to-radius ration $h/R$ only weakly influences this critical sheet thickness where capillary recession leads to truncation.} Figure \ref{fig:MenisqueEpaisseurComp}(c) compares the measured thickness for both Water and UCON$^{\mbox{{\tiny TM}}}$/Water mixture at a $h/R = 0.2$ and a wheel speed of $115$ rpm. {In comparison, the liquid sheet width decreases in a very similar trend for both cases until the point where a liquid rim is formed for the case of water at about $s = 7.5$ cm and $\delta_s \approx 1.5$~cm whereas the formation of a liquid rim occurs much later for the UCON$^{\mbox{{\tiny TM}}}$/Water mixture. In the latter case, the critical sheet thickness $\delta_s$ where recession truncates the sheet is about $0.3$~cm.} This suggests that, compared to the case of water, capillary recession is  delayed for the more viscous UCON$^{\mbox{{\tiny TM}}}$/Water mixture.

	{Finally, it is pointed out that
the contour plots in figure \ref{fig:MenisqueEpaisseur} for the case of $h/R = 0.8$ present a distinct phenomenon. They show that the thickness of the sheet is 
at least as large as the wheel width for all velocities presented here and it is so almost all over the sheet. Also, no characteristic liquid rim is observed when $h/R = 0.8$. This trend is also visible in figure \ref{fig:MenisqueEpaisseurComp}(b) where the liquid sheet thickness (filled, blue diamonds) remains thicker than $4.5$ cm for a good distance away from the wheel and then the sheet suddenly disintegrates without a rim.}

\subsection{Entrained liquid film}
\label{sec-debit}
{As mentioned in the introductory section, the inertial entrainment produced by a rotating wheel does not only eject a thin liquid sheet, but also drags out a liquid film on the wheel rim. We discuss in the following the flow rate entrained in this liquid film. While it is conventional in the LLD approach to characterize the film thickness $\delta_f$, a global flow rate measurement is privileged in this study over film thickness measurements for practical reasons. At the same time, this provides a interesting measurement of the ``efficiency'' of the overall inertial entrainment flux.}

Most of the film flow rate measurements were done using a simple \textit{scraping} technique which consists in applying the sharp edge of a flexible, transparent plastic sheet on the declining side of the wheel rim and thereby scraping the film flow out of the rim. The former is then collected into a large receptacle, including droplets ejected from the film itself for larger rotation speeds. Since disc speed $U$ can show large variations when one scrapes-off the lubrication film, special care was taken to avoid such variations by systematically monitoring the tachometer over the time interval during which the liquid is allowed to flow into the receptacle. This technique showed very good repeatability. It is also possible to use a \textit{local} measurement technique which measures the film thickness at a given location on the cylinder, as is common at low Reynolds number entrainment flows. However, due to the large Reynolds numbers at the scale of the liquid film, a series of \textit{local} film thickness measurements using \textit{Chromatic Confocal Imaging} showed strong spatio-temporal fluctuations of the film thickness at any given point on the rim. In fact, the \textit{scraping} technique is found to be more robust compared to the \textit{local} measurement technique as it directly provides a spatio-temporal average of the film flow rate, during a fixed time interval. 
{This time interval was of 40 seconds for the lowest wheel velocities, but had to be reduced down to 10 seconds for larger velocities, in order to ensure that the volume of the removed sample does not exceed 2 liters, and that the liquid level in the tank remains constant during the measurement.}

\begin{figure}
\begin{center}
\epsfig{file=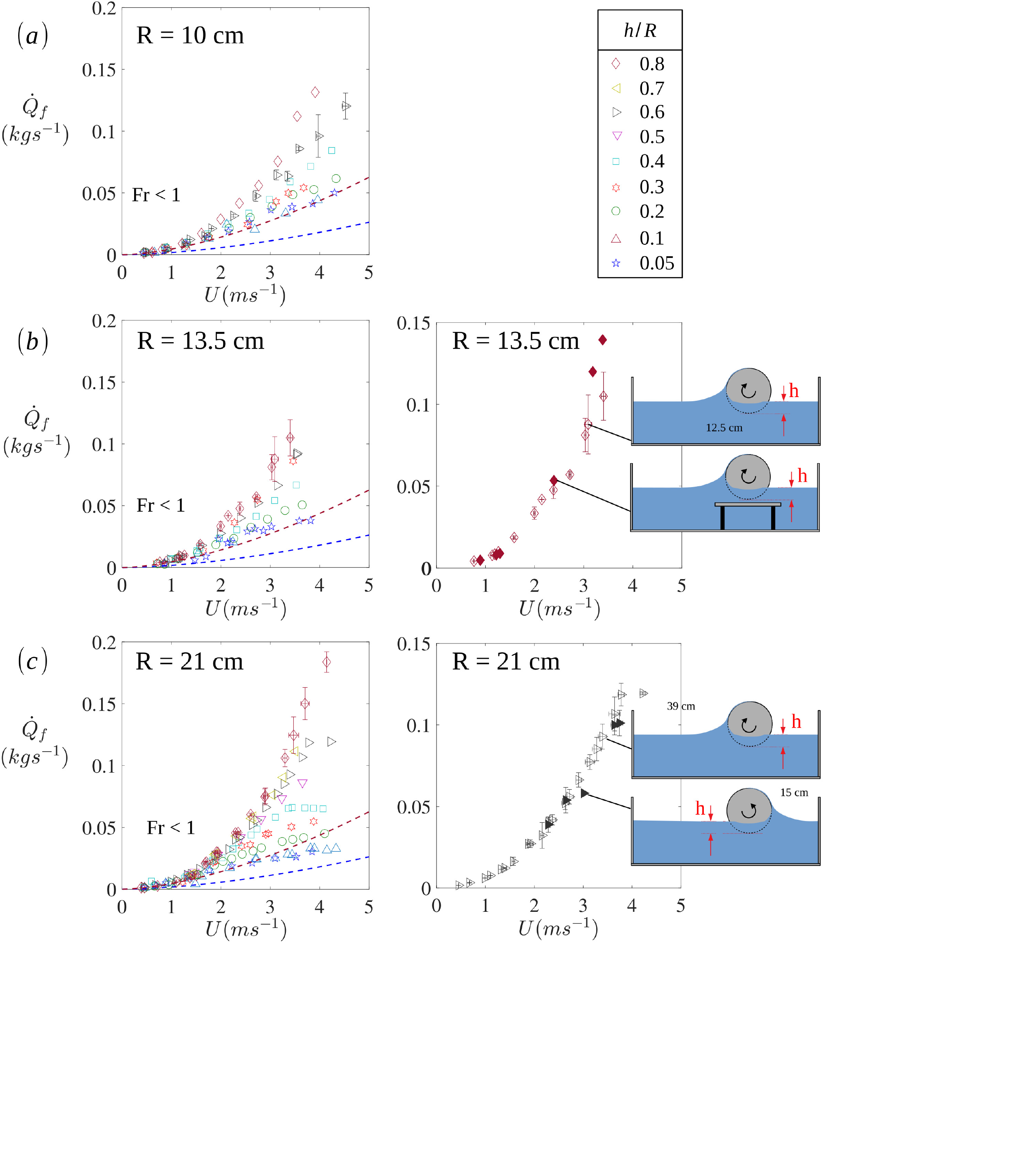, clip, trim=0cm 5cm 5cm 0cm,width=0.9\textwidth,keepaspectratio=true}
\end{center}
\caption{Time-averaged water flow rate  $\dot{Q}_f$ in the entrained liquid film on the rim of the rotating disc. Dashed blue and red lines represent estimations using \ref{eqn:DebitLLDTharma}, as provided by \cite{tharmalingam1978coating}, for respectively the smallest and the largest depth ($h/R = 0.05$ \& $h/R = 0.8$). For the sake of clarity, error bars are given only for one set of data per wheel.}
\label{fig:DebitBrut}

\end{figure}
\begin{figure}
\begin{center}
\epsfig{file=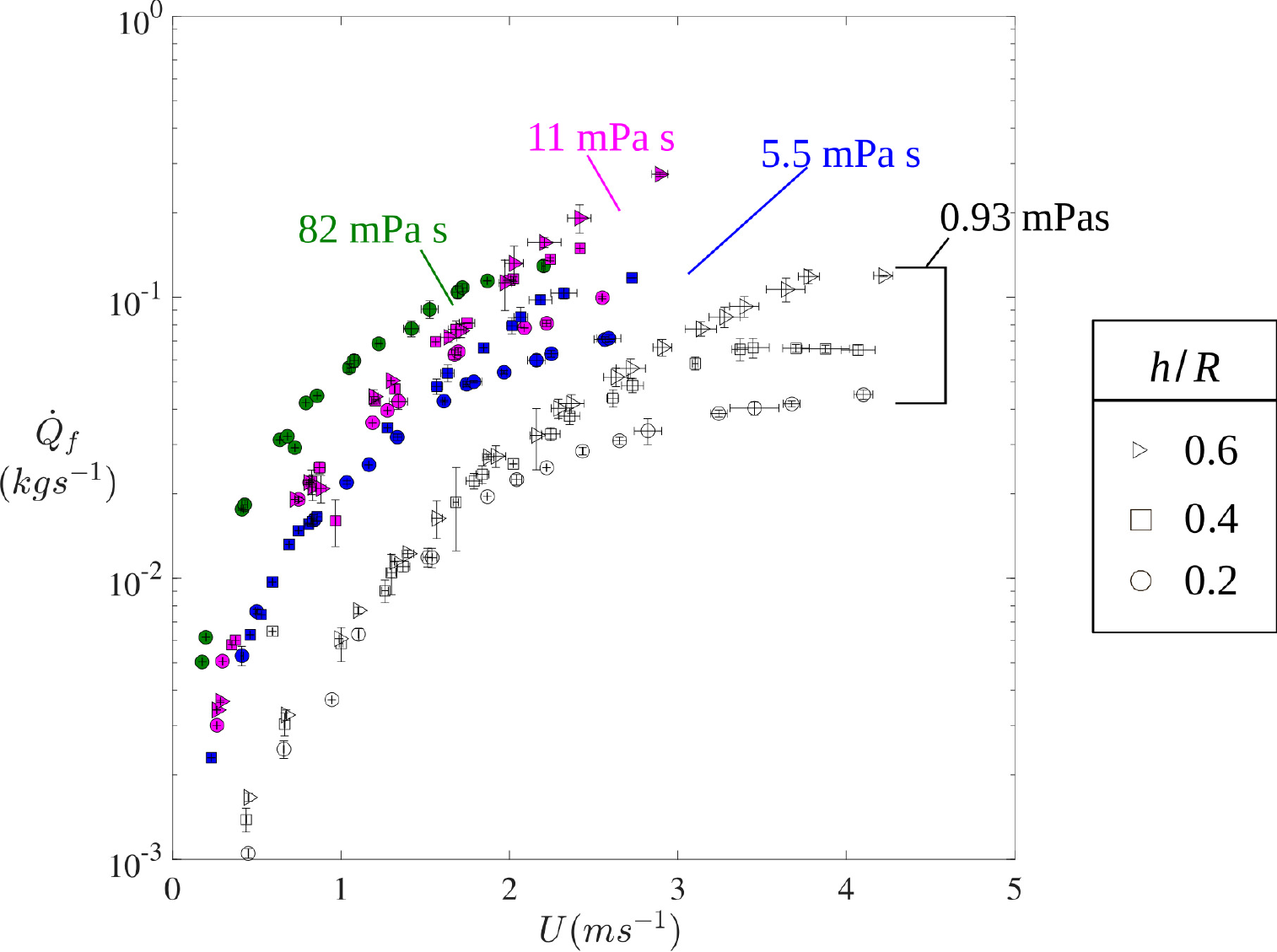,width=0.75\textwidth,keepaspectratio=true}
\end{center}
\caption{Impact of viscosity on the entrained flow rate at a fixed disc radius ($R = 21$ cm).}
\label{fig:DebitBrutVisco}
\end{figure}

Figure \ref{fig:DebitBrut} depicts the variations of the time-averaged film-flow rate as a function of velocity, measured for the three discs of radius $R = 10$ cm, $13.5$ cm and $21$ cm over {at least seven} different immersion heights in water. {Data in figure \ref{fig:DebitBrut} are averaged over different experimental trials for which the rotational speeds differ by at most $5$\%. The error bars in this figure represent the standard deviation of measurements in such samples of trials. For the sake of clarity, figure \ref{fig:DebitBrut} presents these error bars only for a few cases.} It shows that the entrained flow rate increases monotonically with the disc speed $U = R \Omega$ and also with the depth $h$. Note that all data seem to be relatively independent of the water height $h/R$ up to some critical azimuthal speed $U$. After this limit, figure \ref{fig:DebitBrut} clearly indicates that the entrained water flow rate is strongly influenced by the immersion height $h$. For example, the film-flow rate at $h/R = 0.8$ is three times larger than that at $h/R = 0.05$ for discs $R = 10$ cm \& $13.5$ cm. And it gets as large as eight-folds when $R = 21$ cm. 

{It is well-known in dip-coating flows that confinement can play an important role \citep{kim2017confinement} if the meniscus, or the liquid sheet, at the emerging side of the wheel's rim is confined. However, we observe no major effect of confinement in our experiments. We checked this (i) by adding a plexiglass floor below the wheel,  which reduces the distance between wheel and bottom wall from 12.5 cm down to 3 mm (figure \ref{fig:DebitBrut}b) and (ii) reversing the wheel's direction of rotation (the distance between the wheel and the wall goes from {39 cm to 15 cm}, as in figure \ref{fig:DebitBrut}c). In both cases, we could verify that this had no measurable impact on the entrained flow rate.}

The role of the working liquid is explored using the same various water-UCON$^{\mbox{{\tiny TM}}}$ mixtures as in the previous section. Figure \ref{fig:DebitBrutVisco} illustrates that, as expected, increasing the liquid viscosity increases the flow rate. Thus, when the liquid viscosity is increased about $100$-fold, an increase in the liquid entrainment by an order of magnitude is observed. The effect of liquid depth observed for water is also observed for these liquids and, more importantly, figure \ref{fig:DebitBrutVisco} suggests that the corresponding critical velocity depends on the working liquid. 

\subsubsection{Comparison with $2D$ {creeping} flow models}
It is conventional in film coating flows to study the entrained flow by monitoring the film thickness $\delta_f$ as a function of the capillary number $Ca = \mu U / \sigma$. In the creeping flow regime ($Re \ll 1$), the entrained mass flow rate is then given by
\begin{equation}
	\dot{Q}_f = \rho U \delta_f w \left[ 1 - \dfrac{1}{3} \left(\dfrac{ \delta_f}{\delta_f^{g}} \right)^2 \right],
	\label{eqn:Debit}
\end{equation}
where $w$ is the rim thickness {\citep{deryagin1964film, groenveld1970dip}}. Note that the film flow thickness $\delta_f$, at sufficiently small $Ca$, should correspond to the expression  (\ref{eqn:LLDsclaing}) associated with the classical \textit{Landau-Levich-Deryagin} dip-coating flow. However, when $Ca \gg 1$, previous observations \citep{ruschak1985, Kizito1999} report that the observed mass flow rate in the entrained film is obtained from the above expression if the film thickness $\delta_f$ is taken as in the expression (\ref{eqn:DeltaPoidsVisco}). This corresponds to the \textit{viscosity-gravity driven} dip-coating flow and in this case, $\dot{Q}_f = 2/3 \times \rho U w \delta_f^{g}$. More recently, \cite{JinAcrivosMunchPoF2005} suggested, via numerical simulations of creeping flow ($Re \ll 1$) at sufficiently large capillary numbers, that the right expression for the flow rate with $\delta_f = \delta_f^{g}$ should contain a different pre-factor $0.58$ instead of $2/3$. In order to compare our experimental data with these results, we render the experimental flow rates of figure \ref{fig:DebitBrutVisco} dimensionless with  $\dot{Q}_{LLD}$ and $\dot{Q}_{G}$: $\dot{Q}_{LLD}$ is obtained from expression (\ref{eqn:Debit}) when the film thickness is taken to be $\delta_f^{LLD}$ (\ref{eqn:LLDsclaing}), whereas $\dot{Q}_{G}$ is obtained from the same expression (\ref{eqn:Debit}) but with a film thickness of $\delta_f^{g}$ (\ref{eqn:DeltaPoidsVisco}) so that $\dot{Q}_{G} = 0.58 \times \rho U w \delta_f^{g}$.
Figure \ref{fig:Q_Ca} indicates that both $\dot{Q}_{LLD}$ and $\dot{Q}_{G}$ capture the correct order of magnitude of the entrained mass flow rate, which, as inferred from figure \ref{fig:DebitBrutVisco}, varies over more than two orders of magnitude. The main dispersion observed in the rescaled experimental data is due to depth-to-radius ratio $h/R$. In addition, figures \ref{fig:Q_Ca} suggest that this dispersion kicks-in at {some} capillary number which depends on the working liquid's {Morton number}. Finally, below this {characteristic} capillary number, the ratio $\dot{Q}_f / \dot{Q}_{LLD}$ is approximately equal to unity for all liquids while the ratio $\dot{Q}_f / \dot{Q}_{G}$ significantly departs from unity for experiments in water. 

\begin{figure}
\begin{center}
\epsfig{file=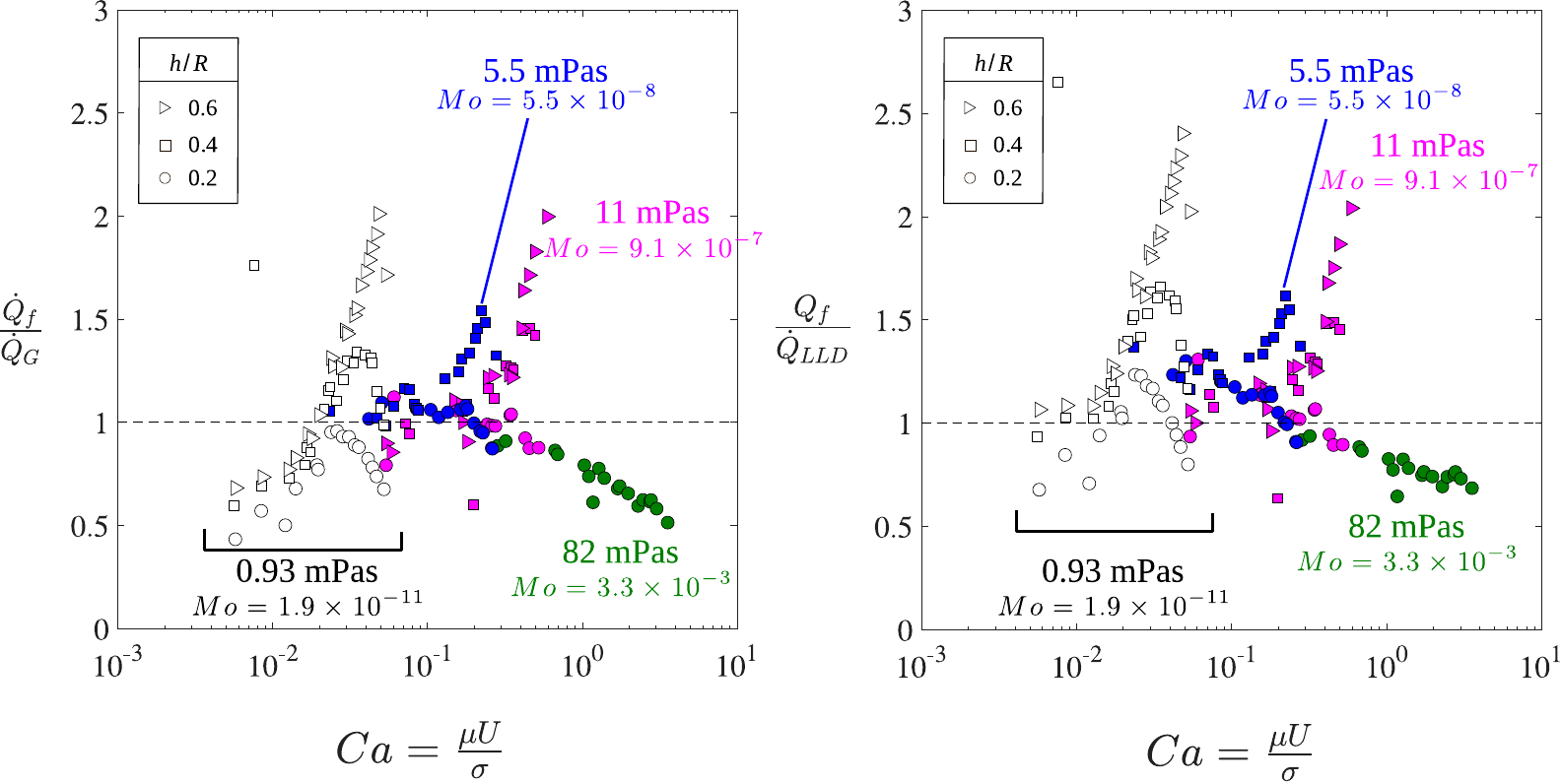,width=1\textwidth,keepaspectratio=true}
\end{center}
\caption{Rescaled data from figure \ref{fig:DebitBrutVisco}. Here, $\dot{Q}_{G}$ and $\dot{Q}_{LLD}$ are obtained from the expression (\ref{eqn:Debit}) when the film thickness $\delta_f$ is taken as $\delta_f^{(g)}$ (\ref{eqn:DeltaPoidsVisco}) and $\delta_f^{LLD}$ (\ref{eqn:LLDsclaing}), respectively.}
\label{fig:Q_Ca}
\end{figure}

\begin{figure}
\begin{center}
\epsfig{file=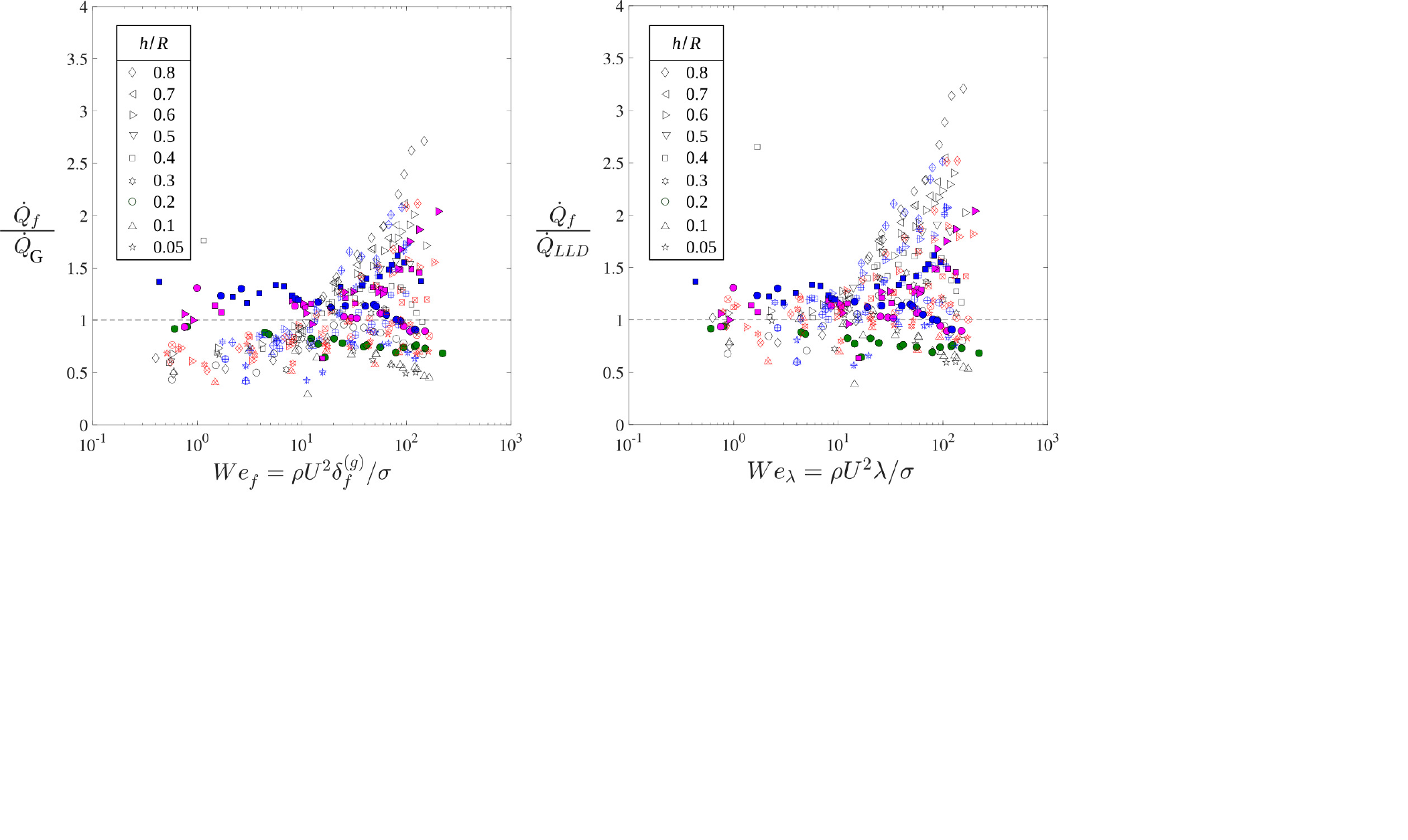, clip, trim=0cm 5cm 5cm 0cm,width=1\textwidth,keepaspectratio=true}
\end{center}
\caption{{Rescaled data from figures \ref{fig:DebitBrut} \& \ref{fig:DebitBrutVisco} as a function of Weber number $We_{f} = \rho U^2 \delta_f^{(g)} / \sigma$ \textit{(left)} and $We_{\lambda} = \rho U^2 \lambda / \sigma$ \textit{(right)} based on the LLD dynamic meniscus length $\lambda = 0.65 l_c Ca^{1/3}$, where $l_c = \sqrt{\sigma / \rho g}$ is the capillary length. Here, open symbols with $+$ (red online) and $\times$ (blue online) represent experimental results for $R = 10$ cm and $R = 13.5$ cm, respectively.}}
\label{fig:DebitRescalingTOUT}
\end{figure}
In fact, it is already known in coating flows
that a transition from the LLD law to an inertia-dominated regime happens when the capillary pressure at the base of the liquid film is no longer large compared to the wall-imposed dynamic pressure $\rho U^2$. {This is precisely the case as the Weber number becomes sufficiently large. Note that two Weber numbers can be defined, depending on the length scale chosen, namely, $We_f$ based on the drainage length scale $\delta_f^{(g)} = l_c  Ca^{1/2}$ \citep{Kizito1999} and $We_\lambda$ based on the quasi-static \textit{LLD meniscus} length $\lambda \simeq 0.65 l_c  Ca^{1/3}$ \citep{de_ryck_quere_1996}. 
We further investigate how the liquid depth modifies the film thickness by plotting all the data from figures \ref{fig:DebitBrut} \& \ref{fig:DebitBrutVisco} as a function of these Weber numbers in figure \ref{fig:DebitRescalingTOUT}.}

{There is a relatively good collapse of data for Weber numbers smaller than $10$, in particular when these data are plotted as a function of $We_\lambda$. In addition, the effect of wheel depth $h$ which kicks in at strikingly different capillary numbers in figure \ref{fig:Q_Ca} (ranging from $10^{-2}$ to $1$) appears above a single threshold Weber number of about $10$, both for $We_\lambda$ and $We_f$. To the authors' knowledge, this is the first ever demonstration of the existence of such a characteristic Weber number above which a transition occurs from  $2D$ creeping flow to an inertial regime for rotary film entrainment. Beyond this Weber number the film thickness is either under-estimated or over-estimated by both the LLD dip-coating law and drainage film length, depending essentially on the ratio $h/R$.}

Note that the rotary flow entrainment problem in our case is particularly distinct from that of classical coating flow processes. Here, instead of a dynamic LLD meniscus at the base of the liquid film flow, a {large, corrugated liquid sheet is present due to inertial ejection of the liquid at the emerging side of the disc's rim}. As already illustrated in section \ref{sec-menisci}, the latter presents a three-dimensional flow structure and its height is proportional to the square of the Froude number. Moreover, the range of Reynolds numbers $Re_f = \rho U_f \delta_f^g / \mu$ in the film flow varies between $\mathcal{O}(1)$ to $\mathcal{O}(10^{3})$. Therefore, it is quite unexpected that a good agreement with $2$D creeping flow analyses  is retrieved here for the entrained flux on a rotating disc, as long as $We_{\lambda} < 10$. In addition, this model provides a satisfactory order of magnitude for the film flow rate even when $We_{\lambda} > 10$.

%

\subsubsection{{Impact of depth on entrained flow rate when $We_{\lambda} > 10$}}

Figures \ref{fig:DebitBrut} (a)--(c) \& \ref{fig:DebitBrutVisco} suggest that beyond a critical speed the role of depth must be accounted for. In this context, \cite{tallmadge1971theory} and \cite{tharmalingam1978coating} suggested a suitably modified version of the dip-coating flow (see schematic \ref{fig:LLDschematics}c). In fact, via lubrication approximation along with LLD-type asymptotic analysis, these authors previously predicted that the average mass flow rate should be
\begin{equation}
\dot{Q}_{TT} = \rho U \delta_f w \left[ 1 - \dfrac{1}{3} \left(\dfrac{ \delta_f}{\delta_f^{g}}\right)^2  \sin \alpha \right],
\label{eqn:DebitLLDTharma}
\end{equation}
where $\delta_f$ is computed from expression (\ref{eqn:LLDTharmaScaling}) and $\alpha$ is the angle between the tangent to the partially immersed cylindrical drum at its line of contact with the liquid bath at rest and the horizontal line. {When the measured data is compared with the estimates from this model (dashed blue and red lines in figures \ref{fig:DebitBrut}), a reasonable qualitative agreement is observed, showing a strong relative increase in the flow rate with increasing depth. This arises from the fact that the model accounts for the squeezing of the meniscus occurring at large $\alpha$, or low depths, and therefore predicts a smaller flow rate than that of equation (\ref{eqn:LLDsclaing}) corresponding to the drag-out problem of a vertical plate ($\alpha = 90^{\circ}$). Nevertheless the model largely underestimates the entrained flow rate. Also, $\dot{Q}_{TT}$ depends on $h/R$ irrespective of the wheel speed $U$ while figures \ref{fig:DebitBrut}(a)-(c) and \ref{fig:DebitRescalingTOUT} suggests that, for all $We_\lambda < 10$, the entrained flux is well predicted by the classical $2D$ creeping flow models corresponding to the vertical drag-out problem. Therefore, the observed influence of depth $h$ when inertia dominates both surface tension and viscous forces (\i.e., at large $Re_f$ and $We_{\lambda}$) cannot be explained by simply accounting for the angle between the emerging side of the disc and the horizontal pool level in the $LLD$ flows.} 

{A first possible explanation can be found in the observation that there exists a thin liquid film which is dragged out of the liquid bath by the disc's lateral walls and which could, in turn, contribute to the film flow on the disc's rim provided that centrifugal forces overcome gravity, which occurs for $Fr>1$.} This contribution is expected to increase when the depth is increased and a larger fraction of the lateral wall is consequently wet. In order to better quantify this contribution, two wiper blades were used to scrap-off the lateral film: the wiper blades were placed right next where the disc's lateral walls and the liquid bath meet. The results are shown in figure \ref{fig:raclage} for $R = 21$ cm at various water depth $h/R$ (filled symbols) where they are also compared with data from figure \ref{fig:DebitBrut}(c). Here, it is evident that the water flow on the lateral wall modifies the entrained flow on the rim. Clearly, data from experiments with lateral scrapping for the case of $h/R = 0.2$ and $h/R = 0.4$ closely follow a unique curve given by $\dot{Q}_{LLD}$. Nonetheless, at water depths corresponding to $h/R = 0.6$ and $0.8$, the supplementary entrainment via lateral walls accounts only for about $10$--$20$ \% increase in film flow rate on the disc's rim. Furthermore, experimental data in figure \ref{fig:raclage} (right) strongly suggests that, for both $h/R = 0.6$ \& $h/R = 0.8$, the entrained flow rate in the absence of contribution from the film flow along lateral walls increases as a power-law of $We_{\lambda}$, independent of $h/R$.

{The physical mechanism for this observed entrained flux augmentation at depth-to-radius ratio $h/R > 0.4$ can be linked with the shape and dynamics of the liquid sheet on the emerging side of the wheel. As illustrated via photographs \ref{fig:MenisqueEAU_R21} and time-averaged sheet thickness measurements \ref{fig:MenisqueEpaisseur} in section \ref{sec-menisci}, when the water depth is sufficiently large, the whole liquid sheet is as thick as the wheel rim whereas the sheet thickness is negligibly small compared to the wheel width when $h/R \leq 0.4$. In the latter case, it is deduced that the liquid sheet does not influence the flow rate at all speeds, whatever the working Weber number $We_\lambda$, as depicted by the corresponding data in figure \ref{fig:raclage} (right) in the absence of any contribution from lateral wall entrainment. However, in the former case, \i.e., when $h/R > 0.4$, the liquid sheet wets the entire thickness of rotating wheel: liquid is entrained from this liquid rim onto the wheel. Taking this action of the liquid sheet into account, it is possible to extend the classical Landau-Levich scaling to explain the liquid entrainment without lateral wall contributions.}

\begin{figure}
\begin{center}
\epsfig{file=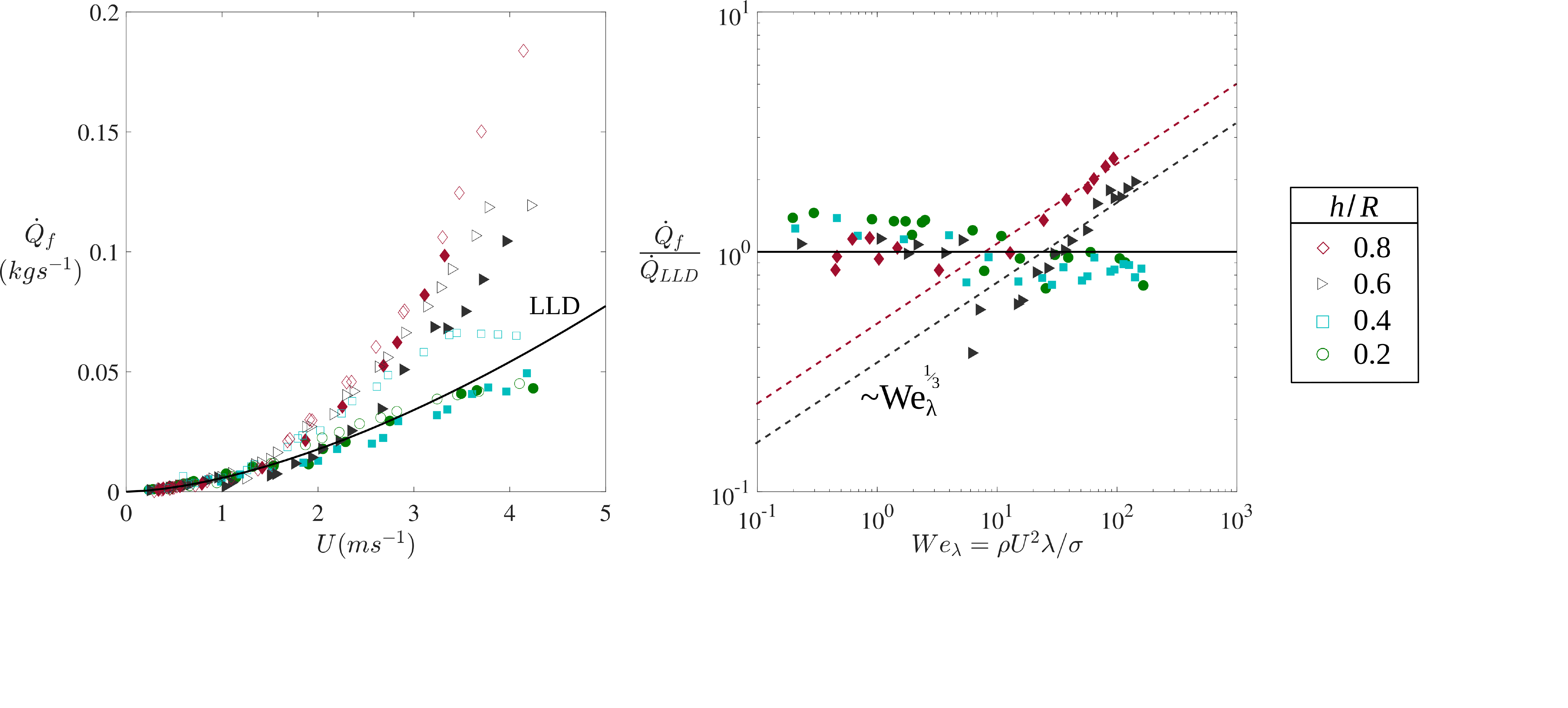, clip, trim=0cm 5cm 5cm 0cm,width=1\textwidth,keepaspectratio=true}
\end{center}	
\caption{Direct evidence for the contribution from liquid entrained via lateral walls (water, and wheel radius $R = 21$ c m). Open symbols represent the cases without lateral scrapping while closed symbols represent data from scrapped cases. Remarkably, data corresponding to the depth to radius ratio $h/R = 0.4$ (black, filled squares) follows the LLD-scaling when lateral entrainment is scrapped off. However, for $h/R = 0.6$ and $0.8$, lateral entrainment cannot completely explain the increase in mass flow rate.}
\label{fig:raclage}
\end{figure}
In the classical entrainment problem, as illustrated by \citep{maleki2011landau}, Landau-Levich asserted that the viscous driving force per unit volume $\mu U/\delta_f^2$ on a fluid element is balanced out by the restoring force per unit volume $(\sigma / l_c)/ \lambda$ which is taken solely due to the pressure gradient arising from the Laplace pressure across the dynamic meniscus of length $\lambda$. Here, $\delta_f$ is the entrained film thickness and $l_c = \sqrt{\sigma/\rho g}$ is the capillary length. The missing length scale $\lambda$ is then obtained by matching the curvature of the dynamic meniscus $\delta_f / \lambda^2$ to the static curvature $1/l_c$. Thereby, it is possible to obtain $\delta_f^{(LLD)}  \sim 	l_c {Ca}^{2/3}$
which is exactly the first-order approximation to the Wilson formula in equation \ref{eqn:LLDsclaing}. However, the fact that the wheel rim is entirely covered by a \textit{thick} liquid sheet from which the thin liquid film emanates is indeed very distinct from the Landau-Levich case wherein the liquid film emerges from a quasi-static liquid bath. Therefore, it is no longer appropriate to match the curvature of dynamic menisci $\delta_f / \lambda^2$ with that of a static menisci $1/l_c$. Since the sheet height $H_m$ depends only weakly on the wheel radius $R$, a simple dimensional analysis points out that the appropriate curvature before the dynamic menisci should be $g/U^2 \times f(h/R)$. By matching the curvature of dynamic menisci $\delta_f / \lambda^2$ with this \textit{inertial} menisci curvature, the film thickness $\delta_f^{I}$ is obtained as 
\begin{eqnarray}
\delta_f^{I}  &\sim 	&l_c {Ca}^{2/3} \left(\dfrac{U^2}{g l_c} \right)^{1/3} \zeta\left(\dfrac{h}{R} \right),
\label{eqn:inertialMalekiLLDscaling}
\end{eqnarray}
where $\zeta(h/R)$ is an arbitrary function of only the water depth to radius ratio. Thus, the corresponding entrained film flow rate is given by
\begin{eqnarray}
\dot{Q}_f^{I}  &\sim 	&\dot{Q}_f^{LLD} We_{\lambda}^{1/3} Ca^{-1/9} \zeta\left(\dfrac{h}{R} \right).
\label{eqn:inertialMalekiLLDscalingBis}
\end{eqnarray}
where only the first-order contribution of the thickness (\ref{eqn:LLDsclaing}) in the expression (\ref{eqn:Debit}) is taken. The modified LLD scaling (\ref{eqn:inertialMalekiLLDscalingBis}) can be compared with the experimental data for the case when the liquid sheet occupies the entire rim \i.e., when $h/R \geq 0.6$. As seen in figure \ref{fig:raclage} (right), despite many simplifying assumptions, the above scaling argument matches relatively well with the available data when the contribution from lateral entrainment is scrapped-off for both $h/R = 0.6$ \& $h/R = 0.8$. Therefore, up to a first estimate, the observed entrainment in the inertial regime of a rotating wheel at $h/R\ge 0.6$ should arise from the presence of a thick liquid sheet.

\section{Conclusion}

{The aim of the present work is to illustrate the dominant flow structures in relation to the rotary entrainment of a partially-submerged disc in the absence of any confinement at the point where the disc emerges out of the liquid pool. In particular, we performed experiments in the range of very large Reynolds $Re_f$ and Weber numbers $We = Ca Re_f$ where the relevant Reynolds numbers is taken as $Re_f = \rho U \delta_f^{g}/\mu$ with the drainage length scale $\delta_f^{g} = \sqrt{\mu U/ \rho g}$.}

{Experiments showed that a liquid sheet of finite height is present on the emerging side of the rotating disc when the rotational speed is sufficiently large ($U \gtrsim 0.8$~m~s$^{-1}$). The time-averaged} liquid sheet height $H_m$ is shown to be proportional to the maximum ballistic height $U^2/2g$ attained by a fluid particle. {It is independent of the immersion depth $h/R$, except for the case of $h/R \geq 0.6$ wherein combined immersion depth and the sheet height become larger than the wheel size $h + H_m > R $.} In particular, when the immersion depth is small compared to the radius ($h/R < 0.5$), {time-averaged} sheet thickness measurements indicate that the sheet rapidly thins out until capillary recession inhibits further decrease. Since the capillary recession is delayed in viscous liquids, the sheet height in viscous UCON$^{\mbox{{\tiny TM}}}$/Water mixtures is found to be much larger than those observed in water at the same velocity. {For large depth-to-radius ratio $h/R$, no capillary recession is observed. Instead a thick liquid sheet as large as the rotating wheel's rim appears which in turn can influence the entrained liquid flux over the rim of the rotating disc.}

{Despite $Re_f \gg 1$ ranging over three decades in the present experiments, the liquid entrainment follows remarkably the flow rates predicted by the $2D$ creeping flow models until a characteristic capillary number $Ca = \mu U / \sigma$, or equivalently a characteristic Reynolds number, depending only on the liquid physical properties (since $Re_f = \sqrt{Ca^3/Mo}$ where the Morton number $Mo = \mu^4 g / \rho \sigma^3$).} The transition to an inertia-dominated entrainment regime occurs when the capillary pressure based on the quasi-static \textit{LLD meniscus} length $\lambda \simeq 0.65 l_c  Ca^{1/3}$ becomes lesser than the wall-imposed dynamic pressure $\rho U^2$. Experimental data in the present study demonstrates that such a transition happens when $We_\lambda = \rho U^2 \lambda / \sigma$ is about $10$. Beyond this limit, two major contributions are identified for inertial entrainment : (1) lateral wall entrainment and (2) entrainment from the top of the thick liquid sheet. The former is present when the Froude number based on the wheel radius is larger than unity and at this stage, the liquid film entrained on the lateral wall is centrifuged towards the wheel rim. This seems to be the dominant mechanism when the water depth is small compared to the wheel radius. The latter modifies the curvature of the quasi-static liquid in front of the entrained film and thereby leads to an enhanced film flow rate. This contribution is predominant when the liquid sheet thickness is comparable to that of the wheel's rim ($h/R > 0.5$).

{We also showed, however briefly, that sheets can meander, become corrugated and fragment into droplets. Furthermore, when the disc width is increased multiple sheets can occur. While \textit{internal}, \textit{external} rimming flows and rotary coating flows have received a lot of attention in the last few decades \citep{seiden2011complexity, schweizer2012liquid}, not much is known for the case of a rotating cylinder submerged in liquid pool in the absence of confinement at the emerging side of the cylinder. So, the authors hope that the present work motivates further numerical simulations and experiments on the complex flow structures occurring in inertia-dominated liquid film flows and also, in the more challenging fully $3$D two-phase flow problem of rotary entrainment at large Reynolds numbers and Weber numbers.}

The authors acknowledge St\'{e}phane Martinez and Arthur Buridon for their technical assistance, {H\'{e}l\`{e}ne Scolan} and also the contribution from projects at the Mechanics department in the Universit\'{e} Claude Bernard Lyon$1$ for several measurements. Financial support from $Groupe$ $PSA$ is also acknowledged by the authors.

Declaration of Interests -- The authors report no conflict of interest.
\begin{landscape}
\appendix
\begin{table}
  \begin{center}
\def~{\vphantom{0}}
  \begin{tabular}{c|c|c|c|c|c|c}
         \hline
      Author(s)(year) &Experiment &Liquid(s)  & $Re_f = \dfrac{\rho \Omega R \delta_f^g}{\mu}$ & $Ca = \dfrac{\mu \Omega R }{\sigma}$ &$We_\lambda = \dfrac{\rho \Omega^2 R^2 \lambda}{\sigma}$ &$Fr = \Omega \sqrt{\dfrac{R}{2g}}$\\
         \hline
         \hline
       \cite{yih1960instability} &RD$^\ddagger$ & WG & $36$ -- $2\times10^{3}$ &$0.01$ -- $4$   &$6$ -- $2\times10^{2}$  &$0.7$ -- $2$\\
       
       \cite{groenveld1970high} &RD &WG, GS & $5 \times 10^{-4}$ -- $9$ &$7 \times 10^{-3}$ -- $6$   &$3 \times 10^{-5}$ -- $10$ &$8 \times 10^{-4}$ -- $0.15$\\
       
       \cite{middleman1978} &RD$^\ddagger$ &WG & $0.02$ -- $2\times 10^{2}$ $^{(a)}$ &$7 \times 10^{-4}$ -- $14$   &$5.6 \times 10^{-3}$ -- $46$ $^{(a)}$ &$0.02$ -- $1.1$ $^{(a)}$\\
       
        \cite{tharmalingam1978coating} &RD &WG$^\dagger$ & $6 \times 10^{-5}$ -- $63$ &$10^{-3}$ -- $30$   &$10^{-6}$ -- $40$ &$3 \times 10^{-4}$ -- $0.5$\\
       
       \textbf{Present work (2020)} &RD &WU & $2.6$ -- $3 \times 10^3$ &$5 \times 10^{-3}$   -- $3.6$ &$0.4$ -- $2 \times 10^2$ &$0.09$ -- $3$\\

       \cite{pitts1961flow} &RMC & WG$^\dagger$  & $0.4$ -- $6$ &$0.05$ -- $2$   &$0.1$ -- $1.4$  &$0.2$ -- $0.3$\\
       
       \cite{mill1967formation} &RMC & P, LO$^\dagger$ & $1$ -- $50$ &$0.08$ -- $3$   &$0.2$ -- $30$  & $0.2$ -- $1$\\
       
       \cite{adachi1988coating} &RMC &WG & $0.15$ -- $3 \times 10^{3}$ $^{(b)}$ &$0.01$ -- $10^2$   &$0.025$ -- $10^3$ $^{(b)}$ &$0.02$ -- $2.5$ $^{(b)}$\\
       
       \cite{coyle1990stability} &RMC &WG & ? &$6 \times 10^{-4}$ -- $0.02$   &? &?\\
       
       \cite{rabaud1994dynamiques} &RMC &S & $10^{-3}$ -- $0.1$ $^{(c)}$ &$50$ -- $10^3$  $^{(c)}$ &$0.01$ -- $30$ $^{(c)}$ &$0.01$ -- $0.2$ $^{(c)}$\\
       
       \cite{gaskell1998experimental} &RMC &Oils & $0.08$ -- $20$ &$2 \times 10^{-3}$ -- $0.1$   &$4 \times 10^{-4}$ -- $1$ &$0.01$ -- $0.4$\\
       
       \cite{ascanio2004forward} &RMC &WG & $3 \times 10^{2}$ -- $2 \times 10^{3}$ $^\star$ &$20$ -- $2 \times 10^{2}$   &$5 \times 10^{3}$ -- $5 \times 10^{4}$ $^\star$ &$7$ -- $20$\\
       
       \cite{owens2011misting} &RMC &WG$^\dagger$ & $4 \times 10^{3}$ -- $5 \times 10^{3}$ $^\star$ &$70$ -- $10^{2}$   &$10^{5}$ $^\star$ &$20$\\
       
       \cite{moffatt1977rotating} &\textit{e}RF & GS & $0.04$ -- $0.14$ &$8$ -- $20$   &$0.1$ -- $1$  &$0.07$ -- $0.2$\\
       
       \cite{KovacBalmer1980_hygrocysts} &\textit{e}RF &WG & $0.6$ -- $26$ &$0.3$ -- $4$   &$0.2$ -- $34$ &$0.1$ -- $1$\\       
       
       \cite{preziosi1988run} &\textit{e}RF &S & $10^{-4}$ -- $0.02$ $^{(d)}$ &$2$ -- $34$ $^{(d)}$  &$3 \times 10^{-4}$ -- $0.1$ $^{(d)}$ &$5 \times 10^{-3}$ -- $0.08$ $^{(d)}$\\
       
       \cite{preziosi1988run} &\textit{i}RF &S & $3 \times 10^{-4}$ -- $7$ $^{(e)}$ &$5 \times 10^{-3}$ -- $10^3$  $^{(e)}$ &$2\times10^{-4}$ -- $22$ $^{(e)}$ &$0.01$ -- $0.6$ $^{(e)}$\\
       
       \cite{Melo1993PRE} &\textit{i}RF &S & $0.08$ -- $13$ &$0.7$ -- $22$   &$0.04$ -- $10^2$ &$0.03$ -- $1$\\
       
       \cite{ThoroddsenMahadevan_ExpFluids1997} &\textit{i}RF &WG & $0.7$ -- $1.4 \times 10^3$ &$4 \times 10^{-3}$ -- $46$   &$0.1$ -- $ 10^{3}$ &$0.1$ -- $3$\\       
       
       \cite{de1994quick} &fiber &Water & $28$ -- $1.7 \times 10^{3}$ $^{(f)}$ &$3 \times 10^{-3}$ -- $0.04$   &$0.08$ -- $80$ $^{(f)}$ &-NA-\\
       
       \cite{Kizito1999} &FC &WT, S & $10^{-3}$ -- $20$ $^{(g)}$ &$10^{-3}$   -- $10$ &$2 \times 10^{-5}$ -- $20$ $^{(g)}$ &-NA-\\
       
       \cite{maleki2011landau} &FC &S & $3 \times 10^{-7}$ -- $0.4$ $^{(h)}$ &$2 \times 10^{-5}$   -- $1$ &$6 \times 10^{-11}$ -- $0.02$ $^{(h)}$ &-NA-\\
       
       \cite{JinAcrivosMunchPoF2005} &FC &NS & $0$ -- $32$ $^{(i)}$ &$10^{-3}$   -- $10^2$ &$0$ -- $10^{2}$ $^{(i)}$ &-NA-\\
       
       \cite{filali2013some} &FC &NS & $0.1$ -- $20$ $^{(j)}$ &$0.01$   -- $6$ &$0.1$ -- $40$ $^{(j)}$ &-NA-\\
       
         \hline
  \end{tabular}
  \caption{{Dimensionless numbers as inferred from previous entrainment flow studies :} RD -- Rotating Drum, RMC -- Rotary/Meniscus Coating, \textit{e}/\textit{i}RF --  \textit{external/internal} Rimming Flows, FC -- Flat plate (vertical/inclined) dip Coating. Whenever possible, data presented here are taken from $^{(a)}${\citet[figures $3$, $4$ \& $9$]{middleman1978}}, $^{(b)}${\citet[figure $4$]{adachi1988coating}}, $^{(c)}${\citet[figure $7$]{rabaud1994dynamiques}}, $^{(d)}$ $^{(e)}${\citet[figures $3$ \& $7$]{preziosi1988run}}, $^{(f)}${\citet[figure $4$]{de1994quick}}, $^{(g)}${\citet[figure $3$]{Kizito1999}}, $^{(h)}${\citet[figure $2$]{maleki2011landau}},  $^{(i)}${\citet[figure $5$]{JinAcrivosMunchPoF2005}},$^{(j)}${\citet[table $1$, figures $8$ \& $19$]{filali2013some}};  GS -- Golden syrup, P -- Paraffin, S -- Silicone oil, LO -- Linseed Oil, WG -- Water/Glycerin mixtures, WT -- Winter \& Triton mixtures, WU - Water/UCON mixutures, NS -- Numerical Simulations. Also, $^\dagger${some data were excluded (for example, non-Newtonian liquids)}, $^\ddagger$ {disc's emerging side is confined}, $^\star${data estimated from accepted physical properties at $25^\circ$C}.}

  \label{tab:ExternalCoatingData}
  \end{center}
\end{table}
\end{landscape}
\bibliographystyle{plainnat}
\bibliography{arXivJFM_InertialEntrainment}

\end{document}